\documentclass[traditabstract]{aa}
\usepackage{epsfig}

\usepackage{graphicx,natbib}
\usepackage{natbib,subfigure} 
\usepackage{amsmath,amssymb}
\usepackage[T1]{fontenc}

\usepackage{txfonts}

\usepackage{hyperref}

\hypersetup{
   colorlinks=true,
   citecolor=blue,
   linkcolor=blue,
   urlcolor=black
   }

\usepackage[francais,english]{babel}

\def\mjup{ M_{\rm J}}
\def\mp{ M_\mathrm{p}}
\def\rp{ R_\mathrm{p}}
\def\ms{ M_\star}
\def\rs{ R_\star}
\def\req{R_\mathrm{eq}}
\def\rtrp{ R_\mathrm{tr,p}}
\def\rtrs{ R_\mathrm{tr,\star}}
\def\rpol{R_\mathrm{pol}}
\def\rjup{R_{\rm J}}

\def \rhoc{\rho_\mathrm{c}}
\def\mur{\mu_\mathrm{R}}
\def\murt{\tilde{\mu}_\mathrm{R}}
\def\s{\mathbf{s}}

\def\np{n_\mathrm{p}}
\def\ns{n_\star}

\def\ai{{\alpha_\mathrm{i}}}
\def\aun{{\alpha_1}}
\def\adeu{{\alpha_2}}
\def\atroi{{\alpha_3}}
\def\asum {{\alpha_1+\alpha_2+\alpha_3}}
\def\Order {{\cal O}}

\def\d{\mathrm{d}}
\def\i{\mathrm{i}}
\def\j{\mathrm{j}}
\def\p{\mathrm{p}}

\def\m{\mathrm{m}}

\def\l{\mathrm{l}}

\def\vr{\mathbf{r}}

\newcommand{\eq}[1]{Eq.\,(\ref{#1})}
\newcommand{\sect}[1]{Sect.\,\ref{#1}}
\newcommand{\eqs}[2]{Eqs.\,(\ref{#1}) and (\ref{#2})}

\newcommand{\ind}[2]{\ #1 _{\!\! \mathrm{\ #2}}}

\newcommand{\pdc}[3]{\left. \frac{\partial \!\! \ #1}{\partial \!\! \ #2}\right|_{#3}}

\newcommand{\dd}[2]{\frac{\mathrm{d} \!\! \ #1}{\mathrm{d}\!\! \ #2}}

\titlerunning{Distorted extrasolar planets}
\authorrunning{Leconte et al.}

\begin{document}

\title{Distorted, non-spherical transiting planets:\\impact on the transit depth and on the radius determination.}

\author{J. Leconte\inst{1,4} \and D. Lai\inst{2,4} \and G. Chabrier\inst{1,3,4} 
}

\institute{ \'{E}cole normale sup\'erieure de Lyon, CRAL (CNRS), 46 all\'ee d'Italie, 69007 Lyon,\\ Universit\'e de Lyon, France (jeremy.leconte, chabrier @ens-lyon.fr)
\and
Center for Space Research, Department of Astronomy, Cornell University, Ithaca, NY 14853 (dong@astro.cornell.edu)
\and
School of Physics, University of Exeter, Exeter
\and
KITP, University of Santa Barbara
}

\date{Received 20 February 2009}

\offprints{J. Leconte}


\abstract{
In this paper, we quantify the systematic impact of the non-spherical
shape of transiting planets, due to 
tidal forces and rotation, 
on the observed transit depth. Such a departure from sphericity leads to
a bias in the derivation of the transit radius from the light curve
and affects the comparison with 
planet structure and evolution models which assume spherical
symmetry.  As the tidally deformed 
planet projects
its smallest cross section area during the transit, the measured effective
radius is smaller than the one of the unperturbed spherical planet
(which is the radius predicted by 1D evolution models). This effect
can be corrected by calculating the theoretical \textit{shape} of the
observed planet.

Using a variational method and a simple polytropic assumption for the
gaseous planet structure, we derive simple \textit{analytical}
expressions for the 
ellipsoidal
shape of a fluid object (star or
planet) accounting for both tidal and rotational deformations. We
determine the characteristic polytropic indexes describing the
structures of irradiated close-in planets within the mass range
$0.3\,\mjup<\mp<75\,\mjup$, at different ages, 
by comparing polytropic models 
with the inner density profiles calculated with the full evolution code. Our
calculations yield a 20\% effect on the transit depth, i.e. a 10\%
decrease of the measured radius, for the extreme case of a 1$\mjup$
planet orbiting a Sun-like star at 0.01AU, and the effect can be
larger for smaller mass objects. For the closest planets detected so
far ($\lesssim 0.05$ AU), the effect on the radius is of the order of
1 to 10\% (three times more for the mean density), by no means a negligible effect, enhancing the puzzling
problem of the anomalously large bloated planets. These corrections
must thus be taken into account for a correct determination of the
radius from the transit light curve and when comparing theoretical
models with observations.

Our analytical expressions can be 
easily 
used to calculate these
corrections, due to the non-spherical shape of the planet, on the
observed transit depth and thus to derive the planet's real
equilibrium radius, the one to be used when comparing models with
observations. They can also be used to model ellipsoidal variations of
the stellar flux now detected in the CoRoT and Kepler light curves.
We also derive directly usable analytical expressions for the moment
of inertia and the Love number ($k_2$) of a fluid planet as a function
of its mass and, in case of significant rotation, for its oblateness.  }


\maketitle

\section{Introduction}
\label{sec:intro}

As the 
measurement
of the radii of close-in transiting planets 
continues to gain
in accuracy, providing stringent constraints
on exoplanet theoretical models, any source of errors in the radius
determination must be determined with precision.  
Current 
ground and space-based photometric observations of the host stars of transiting
planets enable us to address new problems. The first direct detection
with Spitzer of the light emitted by the planet \citep{DHL07} opened a
new path to probe the physical properties of the surface and the
atmosphere of transiting exoplanets. Among the first results of the
\textit{Kepler} mission, the detection of ellipsoidal variations of the host
star induced by tidal interaction with a low mass companion has been
claimed \citet{WOS10}.  
More recently, \citet{CW10a,CW10b} showed that light curve analysis
can put direct constraints on the actual shape of transiting planets. They also investigated the impact of the precession of an
oblate object with a non zero obliquity around the orbital axis on the
shape and timing of the transit signal.

These observations 
motivate
us to investigate the deformation of the
planet with respect to a spherical body, because of tidal or
rotational forces. While 
previous
studies focused on the detectability of
the oblateness of a flattened body, we address in the present paper
the more general problem, namely the determination of the
\textit{general shape} of a planet (or star) distorted by both a tidal
and a centrifugal potential, and its impact on the transit depth and
thus on the determination of its correct radius. In order to compute
the ellipsoidal shape (flattening and triaxiality) of a gaseous body,
we derive in \sect{sec:variational} a simple analytical model of the
internal structure of the object \citep{LRS94a}, based on the polytropic
equation of state. The polytropic indexes are calibrated in \sect{sec:n} by 
comparing with 
numerical models describing
the structure of strongly irradiated gaseous planets \citep{LBC09}.
In \sect{sec:shape}, we present directly usable analytical expressions
giving the shape (oblateness and triaxiality) of a distorted planet
(or star) as a function of its mass and polytropic index and compare
our estimates with the numerical method outlined in Appendix \ref{theoryFig} and with the measured values for the major planets of our solar
system. As a by-product, our model yields analytical expression for
the first gravitational moment $J_2$ and the love number $k_2$ of a
self gravitating fluid body. Finally, \sect{sec:transit} quantifies
the effect of the non-sphericity of the planet on the transit depth.

We find that as the planet transits across the stellar disc, we only
see the smaller cross section of its actual ellipsoidal shape so that
the depth of the transit is {\it decreased} with respect to the
expected signal for a spherical object, as discussed by \citet{LML10} in the case of WASP-12\,b. This implies that the radius
inferred from the light curve analysis, derived under the assumption
of spherical planet and star, {\it underestimates} the real
equilibrium radius of the object. This bias needs to be corrected for
a proper comparison with theoretical 1D numerical simulations of the
structure and evolution of extrasolar planets and enhances the actual
discrepancy between theory and observation for the so called "bloated"
planets.


\section{Variational method for compressible ellipsoids}
\label{sec:variational}

In this section, we briefly describe the energy variational method
developed by \citet{LRS93} and \citet{LRS94a} (hereafter LRS1 and
LRS2) to construct general Darwin-Riemann equilibrium models.  In
\sect{sec:general} we briefly summarize the basic assumptions and the
equilibrium relations are derived in \sect{sec:equilibrium}.  More
details about the method in general, as well as the applications to
compact objects, can be found in LRS1 and LRS2 and references to
equations in these papers are denoted with numbers preceded by "I" and
"II", respectively, in the present paper. Solutions to first order in
the deformation will be derived for tidal and rotational deformations
in \sect{sec:td} and \ref{sec:rd}, respectively. 

\subsection{Model description}\label{sec:general}

Consider an isolated, self-gravitating fluid system in steady state.
The system is characterized by conserved global quantities such as
its total mass $M$ and total angular momentum $J$.
The basic idea in our method is to
model our self gravitating system by a limited number of parameters
$x_1,~x_2,...$, in such a way that the total energy
can be written,
\begin{align}
\label{nrj}
E=E(x_1,x_2,\ldots;\, M, J, ...),
\end{align}
An equilibrium configuration is then
determined by extremizing the energy according to
\begin{align}
\pdc{E}{x_\i}{M,J,...}=0,~~~i=1,2,...
\end{align}

An expression like \eq{nrj} can be written down for the total energy of
a binary system (with components of mass $M$ and $M'$).
We adopt a polytropic equation of state between the pressure $P$ and
the mass density $\rho$,
\begin{align}\label{poly}
P=K\rho^{1+1/n}.
\end{align}
This defines the polytropic index $n$ and the ``entropy'' 
$K$ -- both are constant within the object and sufficient 
(with $M$) to describe the mechanical structure of a given object.  
Under the combined effects of centrifugal and tidal forces, the objects
(stars or planets) achieve nonspherical shapes.  We model these shapes
as triaxial ellipsoids of principal axes $(a_1,a_2,a_3)$ and
$(a'_1,a'_2,a'_3)$, respectively. Throughout this paper, unprimed
quantities refer to the component of mass $M$ while primed quantities
refer to the component of mass $M'$. The three directions along which
our principal axes are measured are, respectively, the line connecting
the center of mass of the two components, its normal contained in the
orbital plane and the direction of the orbital angular momentum
vector. In the simple case of coplanar and synchronous rotation, $a_3$
is simply the polar radius and $a_1$ and $a_2$ are the equatorial
radii of the component measured toward its companion and in the
orthogonal direction, respectively.

Specifically, we assume that the surfaces of constant density within
each object can be modeled as \textit{self-similar ellipsoids}. The
geometry is then completely specified by the three principle axes of
the outer surface.  Furthermore, we assume that the density profile
$\rho(m)$ inside each component, where $m$ is the mass interior to an
isodensity surface, is identical to that of a \textit{spherical}
polytrope with the same volume.  The velocity field, ${\bf v}$, of the
fluid is modeled as either uniform rotation (corresponding to the case
of a synchronized binary system), or uniform \textit{vorticity}, ${\bf
  \nabla}\times {\bf v}$, (for nonsynchronized systems).  The
vorticity vector is assumed to be everywhere parallel to the orbital
rotation axis.

For an isolated rotating gaseous sphere, these assumptions are
satisfied \textit{exactly} when the fluid is incompressible
(polytropic index $n=0$), in which case the true equilibrium
configuration is a homogeneous ellipsoid (Chandrasekhar 1969). For a
binary system, our assumptions are strictly valid in the
incompressible limit only if we truncate the tidal interaction at the
quadrupole order. We adopt this quadrupole-order truncation of the
interaction potential in this paper.

Adding up the orbital separation, $r$, our set of unknowns is
$(r,a_1,a_2,a_3,a'_1,a'_2,a'_3)$ or equivalently
$(r,\rhoc,\lambda_1,\lambda_2, \rhoc',\lambda_1',\lambda_2')$ where
$\rhoc$ is the central density and $\lambda_1\equiv
(a_3/a_1)^{2/3},\lambda_2\equiv (a_3/a_2)^{2/3}$ (when no ambiguity
exists or otherwise stated, primed quantities are defined in the same
manner as unprimed ones by simply making the transformation
$x\rightleftarrows x'$ throughout the equations). The total energy can
be written
\begin{align}
E=U+U'+W+W'+T+W_\i,
\end{align}
where
\begin{align}\label{Uint}
U=\int\! \frac{nP}{ \rho}\, \d m = \bar{k}_1K\rho_c^{1/n}M
\end{align}
is the internal energy of component 1 (cf. Eq.\,(I.3.1)), and
\begin{align}\label{Wself}
W=-\frac{3}{5-n}\frac{GM^2}{ R}\bar{f}=- \bar{k}_2GM^{5/3}\rhoc^{1/3} \bar{f}
\end{align}
is the self-gravitational energy of component 1 (cf. Eq.\,(I.4.6)) with,
\begin{align}\label{k1}
\bar{k}_1 \equiv\frac{n(n+1)}{5-n}\,\xi_1|{\theta'}_1|, 
\end{align}
\begin{align}\label{k2}
\bar{k}_2\equiv\frac{3}{5-n}\,\biggl(\frac{4 \pi |{\theta'}_1|}{\xi_1}\biggr)^{1/3}, 
\end{align}
\begin{align}\label{fbar}
\bar{f}(\lambda_1,\lambda_2)\equiv \frac{A_1a_1^2+A_2a_2^2+A_3a_3^2}{2 (a_1a_2a_3)^{2/3}} ,
\end{align}
\begin{align}\label{Ai}
A_\i\equiv a_1a_2a_3\int_0^{\infty}\!\!\frac{du }{\Pi \cdot(a_\mathrm{i}^2+u)}, 
\end{align}
and
\begin{align}
\Pi^2\equiv(a_1^2+u)(a_2^2+u)(a_3^2+u).\nonumber
\end{align}
$\xi_1$ and $\theta_1$ are the classical variables (dimensionless radius and density taken at the surface) used to describe polytropic gaseous spheres (see \citealt{Ch39}) and $G$ is the gravitational constant. $U'$ and $W'$ are similarly defined.
The kinetic energy in the inertial frame reads
\begin{align}\label{totalT}
T=T_\mathrm{s}+T_\mathrm{s}'+T_0 
\end{align}
where the spin kinetic energy of body 1 ($T_\mathrm{s}$) is given by (cf. Eq.\,(I.5.6))
\begin{align} \label{Ts}
T_\mathrm{s}=\frac{1}{2}I(\Lambda^2+\Omega^2)
-\frac{2}{5}\kappa_nMa_1a_2\Lambda\Omega,
\end{align}
with $\Omega$ being the rotational orbital velocity, $\Lambda$ is a measure of the internal rotation rate in the co-rotating frame,
\begin{align}\label{defkappa}
\kappa_n\equiv\frac{5}{3}\,\frac{\int_0^{\xi_1}\theta^n\xi^4\,d\xi}{
\xi_1^4|{\theta'}_1|}
\end{align}
 is a dimensionless coefficient measuring the 
inertia of the body, and
\begin{align}\label{defI}
I=\frac{1}{5}\kappa_n M (a_1^2+a_2^2)
\end{align}
is the moment of inertia with respect to the rotation axis. Tables giving values of the polytropic constants $\bar{k}_1$, $\bar{k}_2$, $\kappa_n$, $\theta'_1$ and $\xi_1$ as a function of $n$ can be found in LRS1 and in \citet{Ch39}. For non synchronous rotation ($\Lambda\neq 0$), our gaseous body is not in the state of solid-body rotation. A rotation rate can thus not have the usual meaning. To have a sense of the angular velocity, one can take the half of the vorticity (${\bf \omega}=\frac{\nabla\times\mathbf{v}}{2}$, where $\mathbf{v}$ is the fluid velocity vector in the inertial frame) as a proxy\footnote{This choice has no impact on the result to first order because i) there is no cross correlation between tidal and centrifugal distortion at this level and ii) the value of $\omega$ only plays a role to compute the rotational distortion for which solid body rotation is ensured because $a_1=a_2$. In this case, the half vorticity reduces to the usual rotation rate and $\omega=\Omega-\Lambda$. To higher order, this simply highlights the absence of a solidly rotating state and the inadequacy of the parametrization by a rotation rate in such cases.}.  $\omega$ is related to $\Lambda$ by
\begin{align}\label{omega}
\omega=\Omega-\frac{a_1^2+a_2^2}{2a_1a_2}\Lambda.
\end{align}
The orbital kinetic energy $T_0$ is simply
\begin{align}\label{T0}
T_0=\frac{1}{2}\frac{MM'}{M+M'}\,\Omega^2r^2.
\end{align}
Finally, the gravitational interaction energy $W_\i$ reads
\begin{align}
W_\i=-\frac{GMM'}{r} 
-&\frac{GM}{ 2r^3}(2I_{11}'-I_{22}'-I_{33}') \nonumber \\
-&\frac{GM'}{ 2r^3}(2I_{11}-I_{22}-I_{33}), \label{interpotential}
\end{align}
with
\begin{align}\label{Itensor}
I_{\i\j}=\frac{1}{5}\kappa_n M a_\i^2\delta_{\i\j}.
\end{align}

\subsection{Equilibrium relations}\label{sec:equilibrium}

We can now derive the set of equilibrium relations 
\begin{align}
\left\{\pdc{E}{x_\i}{M,J,...}=0\right\}_{\i=1,...,7}\nonumber
\end{align}
yielding seven algebraic equations for our seven unknowns $(r,\rhoc,\lambda_1,\lambda_2, \rhoc',\lambda_1',\lambda_2')$. The details of the transformations necessary to express the total energy as a function of the unknowns and conserved quantities alone and to be able to carry out the differentiation are explained in   \S 2.2.1 of LRS2 and just add technical details not needed here. We will thus give only the results.

\noindent Differentiation with respect to $r$ simply yields the modified Kepler's law for the orbital mean motion $\Omega$
\begin{align}\label{kepler}
\Omega^2 = \frac{G(M+M')}{ r^3}\,(1+\Delta+\Delta'),
\end{align}
with
\begin{align}
\Delta \equiv \frac{3}{2}\,\frac{(2I_{11}-I_{22}-I_{33})}{ Mr^2}.
\end{align}

Differentiation with respect to the central density $\rhoc$ yields the \textit{virial relation},
\begin{align}
\frac{3}{ n}U+W+2T_\mathrm{s}=-\frac{GMM'}{ R}g_\mathrm{t}
\end{align}
with the mean radius $R=(a_1a_2a_3)^{1/3}$ and 
\begin{align}\label{g_t}
g_\mathrm{t}\equiv \frac{R}{ Mr^3}\,(2I_{11}-I_{22}-I_{33})
=\frac{2}{3}\,\frac{R}{ r} \Delta.
\end{align}
Using expressions for $U$ and $W$, we get the equilibrium mean radius
\begin{align}\label{radius}
R=R_0 \left[\bar{f}(\lambda_1,\lambda_2)\left(1-2\frac{T_\mathrm{s}}{|W|}\right)
-\left(\frac{5-n}{3}\right)\,\frac{M'}{M}g_\mathrm{t}\right]^{\frac{n}{n-3}},
\end{align}
where $R_0$ is the radius of the \textit{unperturbed} spherical polytrope given by \citep{Ch39}
\begin{align}\label{spherical_Radius}
R_0=\xi_1 \left(\xi_1^2|\theta'_1|\right)^{\frac{n-1}{3-n}}
\left[\frac{(n+1)K}{4\pi G}\right]^{\frac{n}{3-n}}
\left(\frac{M}{4\pi}\right)^{\frac{1-n}{3-n}}.
\end{align}

\noindent Finally, the differentiation with respect to $\lambda_1$ and $\lambda_2$ yields after some algebra (cf. Eqs.\,(I.8.4), (I.8.5) and (I.8.6)):
\begin{align}\label{shape1}
\left\lbrace 
 \left[2+\frac{\Omega^2}{\mur}-2\frac{a_2\Lambda\Omega}{a_1\mur}+\frac{\Lambda^2}{\mur}\right]\,a_1^2
+a_3^2 \right\rbrace \nonumber \\
=\frac{2}{q_n \murt }(a_1^2A_1-a_3^2A_3),\\
\left\lbrace  \left [\frac{\Omega^2}{\mur}-1-\frac{2a_1\Lambda\Omega}{a_2\mur}+\frac{\Lambda^2}{\mur} \right]\,a_2^2
+a_3^2 \right\rbrace \nonumber \\
=\frac{2}{q_n\,\murt}(a_2^2A_2-a_3^2A_3),\label{shape2}
\end{align}
where $$q_n\equiv\kappa_n(1-\frac{n}{5}),$$ $\mur\equiv GM'/r^3$, $\murt\equiv\mur/(\pi G \bar{\rho})$, and $\bar{\rho}\equiv M/(\frac{4}{3}\pi R^3)$ is the mean density of the ellipsoid.

\section{The shape of gaseous bodies}\label{sec:shape}

In this section, we derive analytical expressions for the deformation
- induced either by centrifugal or tidal potential - of a gaseous
body. To test the validity of our assumptions, we compare our
predictions with the measured values for the major planets of our
solar system in \sect{sec:rd}. Since we do not make any assumption
about the masses of the two components, these equations can be used
indifferently to compute the shape of the star or of the planet by
choosing $M=\mp$ and $M'=\ms$ when considering the planet, and vice
versa when focusing on the star.

In general, the set of equations described in the previous section must be solved numerically, but we will study here the first order development of these equations at large orbital separation. This approximation corresponds to neglecting terms of order $\Order(R_0^5/r^5)$, which is consistent with our truncation of the gravitational potential at the quadrupole order and is appropriate to address close-in transiting planetary systems. In practice, this is done by setting $\Delta=\Delta'=0$ and
\begin{align}\label{defalpha}
a_\i \equiv R_0(1+\ai),
\end{align}
in Eqs\,(\ref{kepler}), (\ref{radius}), (\ref{shape1}), (\ref{shape2}) and their primed equivalent and by expanding these equations to first order in $\ai$ ($\Order(R_0^3/r^3)$).
First we derive some general formulae by expanding the integrand in the definition (\ref{Ai}):
%
\begin{align}
A_\i&=R_0^3(1+\asum)\times \nonumber \\
&\times \int_0^\infty\left[1-\frac{R_0^2}{ R_0^2+u}(\asum+2\ai)
\right]\frac{du}{(R_0^2+u)^{5/2}} , \nonumber
\end{align}
which yields
\begin{align}
A_i=\frac{2}{3}+\frac{4}{15}(\asum)-\frac{4}{5}\ai+ \Order(\ai^2). \nonumber
\end{align}
To first order,
\begin{align}
\bar{f}(\lambda_1,\lambda_2) = 1 + \Order(\ai^2),
\end{align}
and
\begin{align}\label{A1A2A3}
A_1a_1^2-A_3a_3^2&=\frac{8}{15}(\alpha_1-\alpha_3)R_0^2,\cr
A_2a_2^2-A_3a_3^2&=\frac{8}{15}(\alpha_2-\alpha_3)R_0^2.\cr
\end{align}
The principal moment of inertia of the body can also be computed and reads
\begin{align}  \label{dimensionlessmoment}
I=&\frac{1}{5}\kappa_n M (a_1^2+a_2^2)\nonumber \\
\approx&\frac{2}{5}\kappa_n M R_0^2\,(1+\alpha_1+\alpha_2) \nonumber \\
=&\frac{2}{5}\kappa_n M R_0^2~~\mathrm{(spherical\ case)}.
\end{align}
The other moments of inertia can be computed by replacing 1 and 2 by the appropriate indices. The dimensionless moment of inertia $\kappa_n$ for different planetary masses, age and stellar irradiation can be found in tables \ref{tab:cond} and \ref{tab:irrad}.

\subsection{Rotational deformation: Maclaurin spheroids}\label{sec:rd}

Our set of equations also allows us to compute the effect of the centrifugal force alone on a slowly rotationg fluid object. To do so, one just has to take the $M'\rightarrow0$ limit in Eqs\,(\ref{kepler}), (\ref{radius}), (\ref{shape1}) and (\ref{shape2}). Therefore, $\Omega$ is a free parameter (the rotation rate of our body) and there is a degeneracy between $\Omega$ and $\Lambda$ that allows us to choose $\Lambda=0$.

We introduce the dimensionless angular
velocity 
\begin{align} \label{omegabar}
{\bar\omega}^2\equiv\frac{\omega^2 R_0^3}{GM}\end{align}
 as a small
parameter of order $\Order(\ai)$ in all expansions.
The volume expansion factor can be calculated using
\begin{align} \label{rotexpansion}
\frac{T_s}{|W|}= \frac{\frac{1}{2}I\omega^2}{\displaystyle \frac{3}{5-n} \frac{GM^2}{ R}}
= \frac{1}{3}q_n {\bar\omega}^2.
\end{align}
We get
\begin{align}\label{mac1}
\frac{a_1a_2a_3}{ R_0^3}-1=\asum=\frac{2n}{3-n}q_n {\bar\omega}^2. 
\end{align}

To the same order of approximation, the two remaining equations are given by \eqs{shape1}{shape2} which yield
\begin{align}
q_n \bar{\omega}^2=\frac{4}{5}(\alpha_{1,2}-\atroi) \label{mac2}
\end{align}
Combinations of \eqs{mac1}{mac2} give the three figure functions
[cf.~Eq.~(A12) of LRS2]
\begin{align}
\aun=\adeu&=\frac{1}{4}\frac{5+n}{3-n}q_n\bar{\omega}^2\nonumber \\
\atroi&=-\frac{1}{2}\frac{5-3n}{3-n}q_n\bar{\omega}^2. \label{rotation}
\end{align}
For this configuration, the usual variables are the oblateness
\begin{align}\label{oblate}
f=\frac{a_1-a_3}{a_1}=\aun-\atroi=\frac{5}{4}q_n \bar{\omega}^2,
\end{align}
and the dimensionless quadrupole moment of the gravitational field $J_2$ given by the theory of figures to first order \citep{ZMT73}
\begin{align}\label{J2}
J_2=\frac{2}{3}f-\frac{ \bar{\omega}^2}{3}=\frac{5q_n-2}{6}\frac{\omega^2 R_0^3}{GM}.
\end{align}
For practical purposes, $R_0$ can be computed to first order by using \eq{defalpha} and reads
\begin{align}
R_0=\sqrt{3\frac{a_1a_2a_3}{a_1+a_2+a_3}}=\sqrt{\frac{3\,R_\mathrm{eq}^2R_\mathrm{pol}}{2\,R_\mathrm{eq}+R_\mathrm{pol}}},
\end{align}
in our geometry, where $R_\mathrm{eq}$ and $R_\mathrm{pol}$ denote the usual equatorial and polar radii. For an incompressible body ($n=0$) we retrieve the usual solution of the theory of planetary figures $f=~5/4~\bar{\omega}^2$ \citep{ZT80}.

\begin{figure}[htbp] 
 \centering
\resizebox{1.\hsize}{!}{\includegraphics{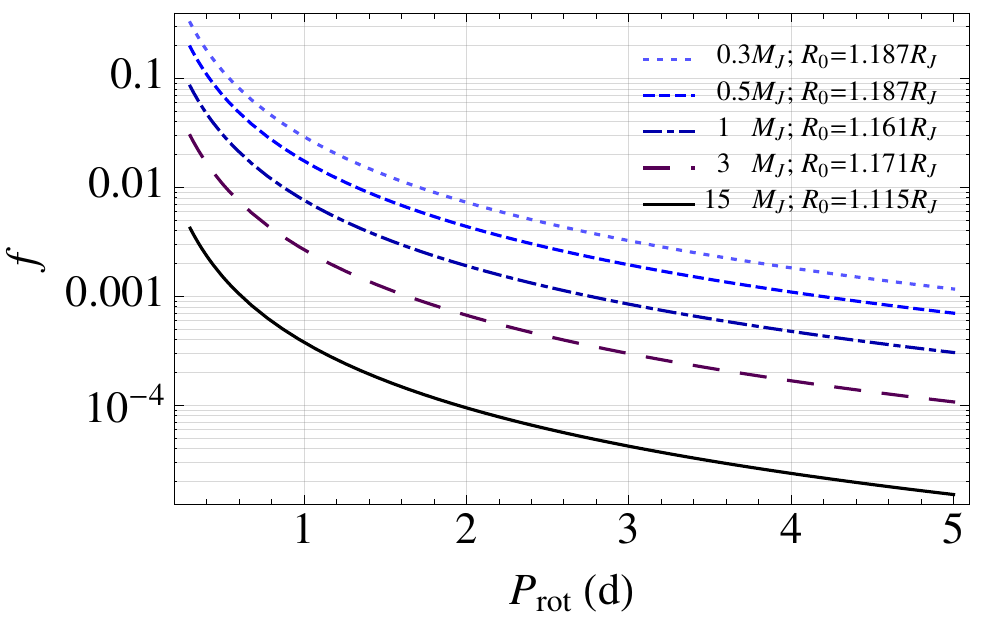}}
 \caption{Oblateness given by \eq{oblate} as a function of the rotation period (in days) at 1 Gyr for planets of mass: 0.3$\mjup$ (dotted), 0.5$\mjup$ (dashed), 1$\mjup$ (dash-dotted), 3$\mjup$ (long dashed), 15$\mjup$(solid). The oblateness decreases when the mass of the planet increases because massive objects are more compressible (see \S4), have a more intense self-gravity field and are thus less subject to perturbations.}
 \label{fig:oblate_irr}
\end{figure}

Attempts have been made to constrain the oblateness and thus the rotation period of transiting planets by using the solar system planets as test cases \citep{CW10a,CW10b}. Because of the wide variety of exoplanets, it is important to have the ability to predict the flattening of fluid planets for a wider range of parameters than encountered in the solar system. Fig.\,\ref{fig:oblate_irr} shows the predicted oblateness for various planet masses as a function of the rotational period $P_\mathrm{rot}=\frac{2\pi}{\omega}$.

\subsection{Tidal deformation:\\determination of the Love number ($k_2$)}\label{sec:td}

To compute the shape induced 
by the tidal force alone, 
we consider a non-rotating configuration ($\omega=0$).
From \eq{omega}, this is achieved if $\Lambda=2a_1a_2\Omega/(a_1^2+a_2^2)$. Then
\begin{align}
T_\mathrm{s}=&\frac{1}{2}I\Omega^2\left(1-\left(\frac{2a_1a_2}{a_1^2+a_2^2}\right)^2\right)\nonumber \\
=&-I\Omega^2\aun\adeu=0+\Order(\ai^2) \nonumber
\end{align}
Thus from \eq{radius} we see that there is
no change of volume to lowest order,
$$\frac{a_1a_2a_3}{R_0^3}-1=0. $$
Since $\Omega^2=\mur(1+p)$ with $p=M/M'$ and $\murt=\Order(R_0^3/r^3)$,
only the zeroth order must be taken in the left hand side of \eqs{shape1}{shape2}, which yields (with help of \eq{A1A2A3})
\begin{align}
\alpha_1-\alpha_3&=\frac{15}{4}q_n\frac{1}{ p}\frac{R_0^3}{ r^3},\cr
\alpha_2-\alpha_3&=0. \cr
 \end{align}
Thus [cf.~Eq.~(A25) of LRS2]
\begin{align}
\alpha_1&= \frac{5}{2}q_n \frac{1}{ p} \frac{R_0^3}{ r^3},\nonumber \\
\alpha_2=\alpha_3&=-\frac{5}{4}q_n \frac{1}{ p} \frac{R_0^3}{ r^3}. \label{tides}
\end{align}
As long as the hydrostatic equilibrium holds, {\it this equation can be used to compute the shape of the planet and its host star at each point of the orbit}. We recover the usual dependence of the tidal deformation in $ \frac{M' R_0^3}{M r^3}$, with a factor of order unity, $q_n$, which encompasses all the structural properties of the gaseous configuration.

Since we are in the linear approximation with a gravitational potential restricted to quadrupolar order, the shape of our body can be described with the usual Love number of second order, $k_2$ (\citealt{Lov09}, which is twice the apsidal motion constant often called $k_2$ in the stellar binary literature). Indeed, once $k_2$ (and $h_2=1+k_2$ for a body in hydrostatic equilibrium) is known, the external potential and the shape that a body will assume in response to any perturbing potential can be computed as detailed in Appendix \ref{theoryFig}. To derive $k_2$, we compute the quadrupolar term of the gravitational potential energy of the system formed by our compressible ellipsoid and a point mass, by introducing \eq{tides} in the linearized version of \eq{interpotential}, and identify this term to the potential energy due to tides given by \citep{Dar08}
\begin{align}
W_\mathrm{tides}=-k_2\frac{G M M' R_0^5}{r^6}.
\end{align}
This yields
\begin{align}\label{lovenumber}
k_2&=\frac{3}{2}\frac{q_n^2}{1-\frac{n}{5}}\nonumber \\
&=\frac{3}{2}\kappa_n^2\,\left(1-\frac{n}{5}\right).
\end{align}
As expected, in the $n=0$ limit, we retrieve the Love number of an incompressible ideal fluid planet $k_2=3/2$. We can also see that $k_2$ is linked to the square of the dimensionless moment of inertia $\kappa_n$. This is because level surfaces are self-similar in our model and that the love number encompass both the deformation of the body ($\propto\kappa_n$) and the gravitational potential created by the deformation ($\propto\kappa_n$). The Love number for different planetary masses, age and stellar irradiation can be found in tables \ref{tab:cond} and \ref{tab:irrad}.

We can see that the value of the Love number tends to decrease with mass above 1$\mjup$. This is due to the fact that more massive objects are more compressible and thus more centrally condensed (See \sect{sec:n}). At constant mass, enrichment in heavy elements toward the center (possibly in a core) acts to decrease the value of $k_2$. In general, redistributing mass from the external to the internal layers, which are less sensitive to the disturbing potential, decreases the response of the body to an exciting potential, which translates into a lower $k_2$.

Our model predicts $k_2$ values in the range 0.3-0.6. As discussed by \citet{RW09} such values of the Love number could be inferred by the measurement of the precession rate of very Hot Jupiters on eccentric orbits. Such measurements could be carried out by \textit{Kepler} for WASP-12\,b analogs with an eccentricity $>3\times 10^{-4}$ (most favorable case) or Tres-3\,b analogs with an eccentricity $>2\times 10^{-3}$ (for $k_2\approx0.3$) and lower eccentricities for higher Love number values. Such measurements would indeed be extremely valuable as they would put direct constraints on the central enrichment in heavy elements inside close Hot Jupiters, like the measurements of the gravitational moments of the solar system planets.

\subsection{Synchronized planets}\label{sec:sp}

For values of the tidal dissipation factors inferred for 
Jupiter \citep{GS66,LCB10}, the timescale of pseudo synchronization of close-in giant planets is less than about a million years. The planet is thus in a state of pseudo synchronization, with a rotation rate given by \citep{Hut81,LCB10}
\begin{align}
\omega_\p= \frac{1+\frac{15}{2}e^2+\frac{45}{8}e^4+\frac{5}{16}e^6}{(1+3e^2+\frac{3}{8}e^4)(1-e^2)^{3/2}}\,\Omega
\end{align}
in the weak friction theory, with
$e$ being the eccentricity of the orbit.
For the simple case of a circular orbit, the spin is thus synchronized and, either solving Eqs\,(\ref{kepler}), (\ref{radius}), (\ref{shape1}) and (\ref{shape2}) in the synchronized case ($\Lambda=0$), or simply adding the results of \eqs{tides}{rotation}  (there is no cross correlation terms to first order) yields
\begin{align}
\alpha_1&=\frac{1}{3}q_n\left(\frac{1+p}{ p}\right) \frac{R_0^3}{ r^3}\left[\frac{5}{4}
          \left(\frac{7+p}{1+p}\right)+\left(\frac{2n}{3-n}\right)\right] \cr
\alpha_2&=-\frac{1}{3}q_n\left(\frac{1+p}{ p}\right) \frac{R_0^3}{ r^3}\left[\frac{5}{4}
          \left(\frac{2-p}{1+p}\right)-\left(\frac{2n}{3-n}\right)\right] \cr
\alpha_3&=-\frac{1}{3}q_n\left(\frac{1+p}{ p}\right) \frac{R_0^3}{ r^3}\left[\frac{5}{4}
          \left(\frac{5+2p}{1+p}\right)-\left(\frac{2n}{3-n}\right)\right], \cr
 \end{align}
and
\begin{align}\label{rotexpansion2}
\frac{R^3}{R_0^3}=1+q_n\left(\frac{2n}{3-n}\right)\,\left(\frac{1+p}{p}\right)\frac{R_0^3}{ r^3}.
\end{align}

\subsection{Model Validation}\label{sec:valid}

There are two major assumptions in the present calculations:
\begin{itemize}
\item The absence of a central core. The aim of such an approximation is to avoid to introduce any free parameter in the model. In any case, the core mass and the global enrichment in giant extrasolar planets are yet weakly constrained \citep{Gui05,LBC09}. We will show that this approximation introduces an uncertainty of $\approx 10\%$ on the derived shape.
\item The polytropic assumption. This allows us to derive a \textit{completely analytical model}. Comparison with a more detailed numerical integration (See Appendix \ref{theoryFig}) shows that the deviation between the results of the two models (analytical vs numerical) is smaller than the uncertainty due to the no-core approximation.
\end{itemize}

Since the oblateness ($f$), $J_2$, $k_2$, the mean radius and rotation rate
are known for the major planets of our solar system, we can test our
theory on these objects. The details of the calculation of the chosen
polytropic indexes are presented in \sect{sec:n}. The results are
summarized in table \ref{tab:moments}, which shows the actual values of
the relevant parameters for the two major planets taken from
\citet{Gui05}, the values of the oblateness and of $J_2$ calculated
with our model and, for comparison, with the assumption of an
incompressible body ($n=0$). 
We see that, whereas the values of $f$ derived from the incompessible
model differ from the true values by almost a factor of 2, our
polytropic model predicts the $f$-values to within $12\%$. 
Note that higher-order terms (of order $\bar\omega^2\sim 10\%$)
are not completely negligible for rapidly rotating bodies such as 
Jupiter and Saturn (see \citealt{ZMT73,CSH92}). The
polytropic model, however, yields the $J_2$-values that differ from
the measured values by $30\%$ (for Jupiter) and $59\%$ (for Saturn).
These discrepancies are mostly due to the large metal enrichment in 
these planet interiors, probably with the presence of a large dense core as detailed three layers models can reproduce exactly the measured moments \citep{CSH92}. 
Note that this no-core approximation has less relative impact on the distortion of the shape predicted by the model than on the gravitational moments because these effects scale as $h_2=1+k_2$ (See Appendix \ref{theoryFig}) and $k_2$ respectively, $k_2$ being $\lesssim 0.6$ in the situations of interest.
Such discontinuities in the density profile (and its derivatives) could 
be addressed more precisely with two different polytropes, but this 
would add extra free parameters and would not serve the very purpose 
of the present paper.

\begin{table}[htb]
\begin{center}
\caption{Comparison between the measured oblateness and data of the gravity fields and the values obtained with our polytropic model (polytrope) and with a model with $n=0$ (incompressible).}
\label{tab:moments}
\small
\begin{tabular}{l r@{.}l r@{.}l } \hline\hline 
	&\multicolumn{2}{c}{\bf Jupiter} &\multicolumn{2}{c}{\bf
Saturn}  \\ \hline
$\mp$ [$10^{26}$kg] & 18&986112(15)  & 5&684640(30)  \\
$R_{\rm eq}$ [$10^7$m] &  7&1492(4)  & 6&0268(4)  \\
$R_{\rm pol}$[$10^7$m] &  6&6854(10)  & 5&4364(10)    \\
$R_0$~~[$10^7$m] &  6&9894  & 5&8198    \\
$P_{\mathrm{rot}}\,$[$10^4$s] & 3&57297(41)  & 3&83577(47)  \\
$\bar{\omega}^2\times10^2$ &  8&332  & 13&940 \\
$k_2$ & 0&49 & 0&32 \\
\hline
$n$ & 0&936& 0&748 \\
$q_n$ &0&547& 0&623 \\
\hline
$f\times10^2$ & 6&487(8)& 9&796(9)   \\
$f\times10^2$ (polytrope) & 5&701  & 10&849   \\
$f\times10^2$ (incompressible) & 10&416 & 17&425   \\
\hline
$J_2\times10^2$ & 1&4697(1)  & 1&6332(10)    \\
$J_2\times10^2$ (polytrope)& 1&023 & 2&586    \\
$J_2\times10^2$ (incompressible)& 4&166 & 6&970    \\

\hline\hline
\multicolumn{5}{c}{
The numbers in parentheses are the}\\
\multicolumn{5}{c}{uncertainty in the last digits 
of the value. }
\end{tabular}
\normalsize
\end{center}
\end{table}

While we decided to use a polytropic assumption to infer a fully analytical theory, the figures of a body in hydrostatic equilibrium can be derived without this assumption. As shown in \citet{Ste39} (See also \citealt{ZMT73,ZT80,CSH92} for more detailed applications to the giant planets case) and outlined in Appendix~\ref{theoryFig}, this theory, however, requires a numerical integration even to first order. 
For an ideal $n=1$ polytropic sphere, \eq{lovenumber} agrees with the numerical results of \citet{Ste39} with less that 1\% error.
To compare these methods in our context, we derive the values of $k_2$ using our analytical model (\eq{lovenumber}), and by numerical integration of \eqs{eqdfeta}{LoveNum} for our best representative models of Jupiter and Saturn in our grid (Although without cores). For Jupiter, our \eq{lovenumber} gives $k_2=0.55$, the numerical integration gives $k_2=0.57$ to be compared with the measured value of $k_\mathrm{2,J}=0.49$. For Saturn, these values are 0.60, 0.66 and $k_\mathrm{2,S}=0.32$ respectively. Both models predict $k_2$ values than are higher than the measured ones. A direct consequence of the presence of heavy elements inside our giant planets. Comparing the models, our \eq{lovenumber} yields slightly smaller $k_2$ values than the numerical integration which tends to mimic a central over-density (See \sect{sec:td}). As discussed above a more precise modeling requires the addition of central enrichment in heavy elements whose mass fraction would be a free parameter. Without better knowledge of the internal composition of giant exoplanets, we think that the two methods yield similar results up to the sought level of accuracy. For sake of completeness, the values of $k_2$ computed with both methods are presented in tables \ref{tab:cond} and \ref{tab:irrad}.


\section{Polytropic index in gaseous irradiated planets}\label{sec:n}

To readily use the results of \sect{sec:shape}, one only needs to have
a proper value for the polytropic index $n$ to be used. In this
section, we derive realistic polytropic indices from numerical models
of gaseous irradiated planets. All the other polytropic functions
($\kappa_n$, $q_n$, ...) can be derived by integrating the Lane-Emden equation and
are tabulated in \citet{Ch39} and LRS1. They are given for different planetary masses, age and stellar irradiation in tables \ref{tab:cond} and \ref{tab:irrad}. We focus on the polytropic
index in the planet because, in the context of transiting exoplanets,
both the stellar rotation and the stellar tides have a negligible
impact on the transit depth, as will be discussed in
\sect{sec:transit}.  The main physics inputs (equations of state,
internal composition, irradiated atmosphere models, boundary
conditions) used in the present calculations have been described in
detail in previous papers devoted to the evolution of extrasolar giant
planets \citep{BCB03,CBB04,LBC09}, and will not be repeated here.

\begin{figure}[htbp] 
 \centering
 \resizebox{1.\hsize}{!}{\includegraphics{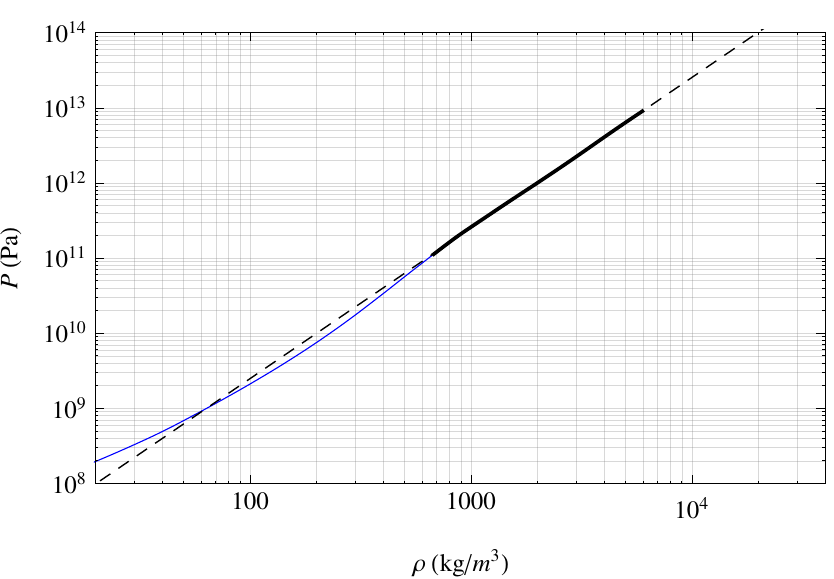}}
 \caption{
The internal pressure-density profile of an irradiated
   1.8$\mjup$ planet (solid line). The dashed line represents
the best-fit polytropic equation of state.
The pressure-density range covered in the inner part of the body (95\%
   in mass) is represented by the thicker part of the solid curve,
   which is well modeled by a polytropic EOS. As the thin part of the
   $P-\rho$ curve represents only 5\% in mass of the body it is
   disregarded by the fit.}
 \label{fig:lrlp-1_8irrad}
\end{figure}

\begin{figure}[htbp] 
 \centering
 \resizebox{1.\hsize}{!}{\includegraphics{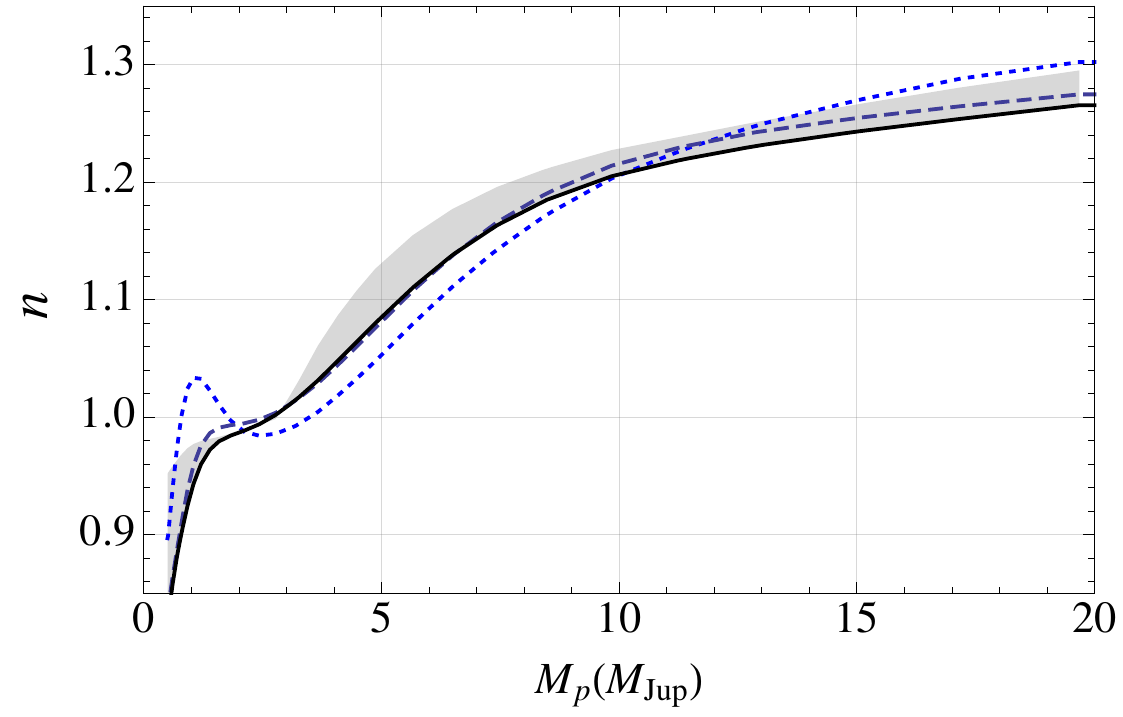}}
 \caption{Polytropic index for
non-irradiated planets
as a function of the planet's mass $\mp$ at 100\,Myr (Dotted), 1\,Gyr
(Dashed) and 5\,Gyr (Solid). The shaded area represents the
uncertainty on the polytropic index for the 5\, Gyr case (see text).}
 \label{fig:cond}
\end{figure}

\begin{figure}[htbp] 
 \centering
 \resizebox{1.\hsize}{!}{\includegraphics{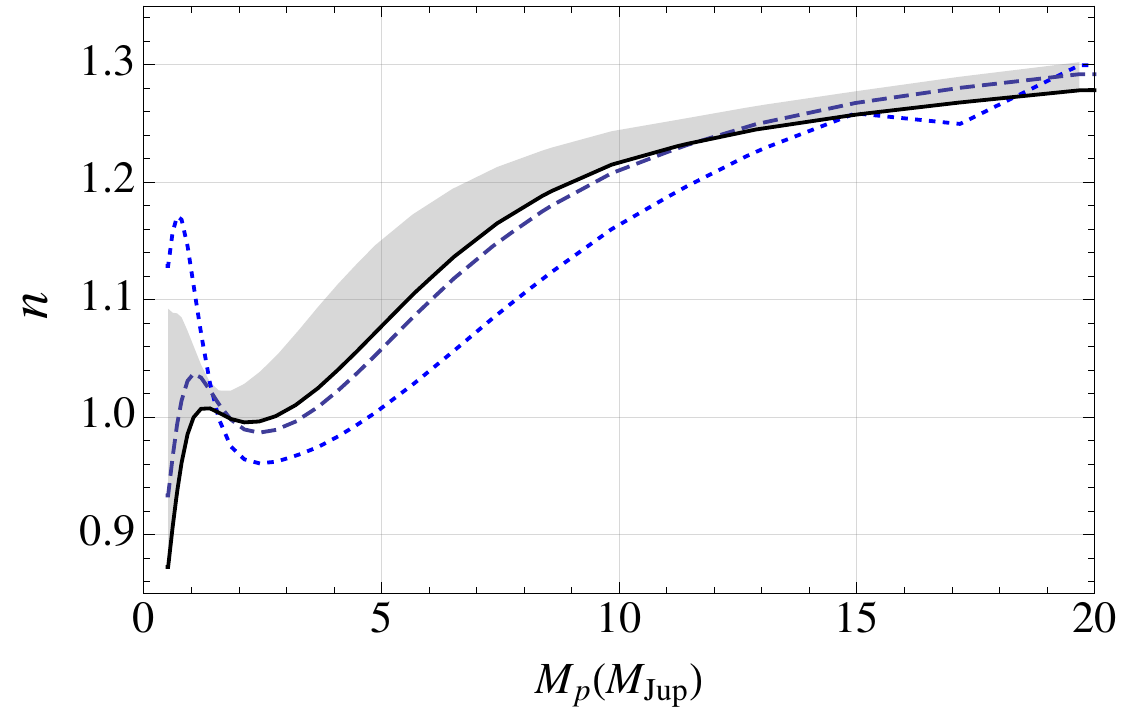}}
 \caption{Polytropic index for strongly-irradiated planets as a
   function of $\mp$ at 100 Myr (dotted), 1 Gyr (dashed) and 5 Gyr
   (solid). As the irradiated atmosphere impedes the radiative cooling
   of the objects, it retards its contraction. Therefore, the
   non-monotonous behavior observed at the early ages in the
   non-irradiated case (Fig.\, \ref{fig:cond}) is enhanced, even at a
   later epoch. The bump at the high mass end of the 100 Myr curve is
   caused by deuterium burning (see text). The shaded area represents
   the uncertainty on the polytropic index for the 5\, Gyr case (see
   text).}
 \label{fig:tr132}
\end{figure}

We computed a grid of evolution models of gaseous giant planets with
solar composition for various masses $\mp\in[0.35\mjup,\,20\mjup]$ and
incoming stellar flux
$F_\star~\in~[0,\,4.18\times~10^6\,\mathrm{W.m^{-2}}]$. Low irradiation
model can be used to infer the oblateness of long period rotating
planets (such as Jupiter), while strongly irradiated models can be
used to infer the shape and its impact on the transit of close-in
planets. For the non irradiated case, the grid extends to $75\mjup$. As the effect of irradiation on the internal structure decreases with the effective temperature of the object, these models computed with non irradiated boundary conditions should give a fair description of massive brown dwarfs in the range of irradiation considered.
The pressure-density profile of each model is then fitted by a
  polytropic equation of state (\eq{poly}) at each time step and an
  example of the result of such a fit is shown in
  Fig.\,\ref{fig:lrlp-1_8irrad}. Note that the disagreement between the actual
  $P-\rho$ profile and the polytrope in the lower left area of
  Fig.\,\ref{fig:lrlp-1_8irrad} is both expected and needed: 
This low-density region (the first 5\% in mass below
the atmospheric boundary surface) has a different effective polytropic
index than the planetary interior. In order to capture the bulk mechnical
property of the planet, 
we weight each shell in the  internal structure profile by its mass 
during the fitting procedure.
This provides us with a grid tabulating the polytropic index of the planet,
$\np \equiv \np(\mp, F_\star, t)$, and its spherical equilibrium
radius, $R_{0,\p}\equiv R_{0,\p}(\mp, F_\star, t)$, where $t$ is the
age of the object. These functions, along with other quantities ($\ind{T}{eff}$, ...), are tabulated in tables \ref{tab:cond} and \ref{tab:irrad}\footnote{Electronic versions of the model grids are available at \textit{http://perso.ens-lyon.fr/jeremy.leconte/JLSite/JLsite /Exoplanets\_Simulations.html}}. Figures \ref {fig:cond} and \ref {fig:tr132} show
the variation of $n$ with the mass for different ages with $F_\star=0$
and $4.18\times~10^6\,\mathrm{W.m^{-2}}$, respectively.

%


As shown in Fig. \ref{fig:cond}, in the non-irradiated case, we
recover qualitatively the results of \citet{CBL09}: except for the
early stages of the evolution, the (dimensionless) isothermal
compressibility of the hydrogen/helium mixture is a monotonically
increasing function of the polytropic index, $\bar{\chi}=\pdc{\ln
  \rho}{\ln P}{T}=\frac{n}{1+n}$, and thus of the mass of the
object. In the high mass regime, $n$ slowly increases as the relative
importance of ionic 
Coulomb
effects compared with the degenerate electron 
pressure
decreases, and approaches the $n=3/2$ limit, the
expected value for a fully degenerate electron gas, when $\mp$
approaches the hydrogen burning minimum mass ($\approx 70\mjup$) as can be seen in table \ref{tab:cond3}. In
the low mass regime, the compressibility decreases with the mass
because the
repulsive Coulomb
potential between the ions, and thus
the ionic electrostatic energy becomes dominant. Ultimately,
electrostatic effects dominate, leading eventually to
$\bar{\chi}\approx n\approx0$ for solid, terrestrial planets.

A new feature highlighted by the present calculations is the non-monotonic behavior occurring between 1 - 3 $\mjup$ at early ages. This occurs when the central regions of the planet, of pressure $P_\mathrm{c}$ and temperature $T_\mathrm{c}$, previously in the atomic/molecular regime, become pressure-ionized, above 1-3 Mbar and 5000-10\,000K \citep{SCV95,CSH92,SHC92}, and the electrons become degenerate. An effect more consequential for the lowest mass objects, whose interiors encompass a larger molecular region. This stems from the fact that \citep{Ch39}
\begin{align}
P_\mathrm{c}>\frac{G\mp^2}{8\pi\rp^4},
~~\mathrm{and}~~\frac{G\mp^2}{8\pi\rjup^4}\approx 2-3 \, \mathrm{Mbar}\approx P_\mathrm{{\rm ionization}}.
\end{align}
Older (with smaller $R_p$)
and more massive ($\mp\gtrsim2\mjup$) objects have
$P_\mathrm{c}>10\,P_\mathrm{{\rm ionization}}$ and the ionization
extends all the way up to the outermost layers of the gaseous
envelope, which then contains too small a mass fraction of molecular
hydrogen to significantly affect the value of the polytropic
index. This contrasts with younger objects around 1 - 3 $\mjup$, whose
external molecular hydrogen envelope contains a significant fraction
of the planet's mass, leading to a larger value of the polytropic
index, as molecular hydrogen is more compressible than ionized
hydrogen (see e.g. Fig. 21 of \citealt{SCV95}). Once again, for these
latter objects, the interior structure would be better described by
using two different polytropes, but such a significant complication of
the calculations is not needed at the presently sought level of
accuracy.

As seen on Fig.\, \ref{fig:tr132}, a strong irradiation enhances the
aforementioned feature: the evolution is delayed because 
the irradiated atmosphere 
impedes the release of the internal
gravothermal energy. This yields a slower contraction, thus a lower
central pressure (and lower central temperature) for a longer period
so that the object enters the ionization regime at a later epoch. The
bump at the high mass end of the 100 Myr isochrone is due to deuterium
burning which also occurs later for a given mass, because of the
cooler central temperature (see above). At 100 Myr, the 20 $\mjup$ has
already burned a significant amount of its deuterium content and
starts contracting again, whereas lower mass planets are still burning
some deuterium supply, leading to a less compact and thus less ionized
structure. This leads to the non-monotonic behaviour on the high-mass
part of the $n-M$ diagram at 100 Myr, which reflects a similar
behaviour in the mass-radius relationship.
%

To evaluate the uncertainty in our determination of the polytropic
index, we use an alternative 
method to derive $n$. As shown by \citet{Ch39}, the knowledge of $M$, $K$ and $n$ is sufficient
to infer the radius of the polytrope, with the help of
\eq{spherical_Radius} and the central density, using
\begin{align}\label{centraldensity}
\ind{\rho}{c}^{\frac{3-n}{2n}}=\frac{M}{4\pi}\left(\frac{4\pi G}{(n+1) K}\right)^{3/2} \left(\xi_1^2|\theta'_1|\right)^{-1}.
\end{align}
Since our numerical simulations provide both the radius,
$R_{0,\p}(\mp, F_\star, t)$, and the central density of the object,
$\ind{\rho}{c,p}(\mp, F_\star, t)$, we can invert
\eqs{spherical_Radius}{centraldensity} to compute $\ind{K}{p}$ and
$\np$. This new determination of the polytropic index is compared with
the previous one, obtained by fitting the $P-\rho$ profile, in Figs
\ref{fig:cond} and \ref{fig:tr132} for the 5\,Gyr case: the new $\np$
value corresponds to the upper envelope of the shaded area. Fig
\ref{fig:cond} shows that the two approaches yield very similar
results in the non-irradiated case. For the irradiated case, the
average uncertainty on our determination of $\np$ lies between about
5\,\% and 15\,\% for the low mass planets.


\section{Implications for transit measurements}\label{sec:transit}

When limb darkening is ignored, the depth of a transit is given by the
ratio of the planetary and stellar projected areas. When both bodies
are spherical, this simply reduces to $\delta L_\star /L_\star\propto
(\rp/\rs)^2$. For close-in planet-star systems, however, both tidal
and rotational deformations yield a departure from sphericity, so that
what is measured is no longer the mean radius but an
\textit{effective} "transit radius" defined 
such that 
the cross section of the
planet is equal to $\pi \rtrp^2$ and similarly for the star. Thus the
transit depth $\delta$ reads
\begin{align}\label{transitdepth}
\delta\equiv\frac{\delta L_\star}{ L_\star}= \left(\frac{\rtrp}{\rtrs}\right)^2.
\end{align}


\subsection{Impact on transit depth}

In general, the projected area of an ellipsoid can be computed for any orientation and then at each point of the orbit, as explained in Appendix \ref{geometry}. Figure \ref{fig:projec} shows the projected area of the planet ($\pi \rtrp^2$) as a function of its anomaly ($\phi$) and inclination ($i$) 
\begin{align}
\rtrp^2=  \sqrt{a_3^2 \sin ^2i \left(a_1^2 \sin ^2\phi+a_2^2 \cos ^2\phi\right)+a_1^2
   a_2^2 \cos ^2i},
\end{align}
normalized to the spherical case ($\pi R_{0,\p}^2$).

When the planet is seen from its "side" ($\phi/\pi=0.5$), the observer sees a bigger planet because the rotation of the latter on itself tends to increase its volume, as has been mentioned by \citet{LML10} for WASP-12\,b. The possibility to measure these effects from the light curve is discussed in \citet{RW09} and \citet{CW10a}.

\begin{figure}[htbp]
\begin{center}
 \subfigure{ \includegraphics[width=8.5cm]{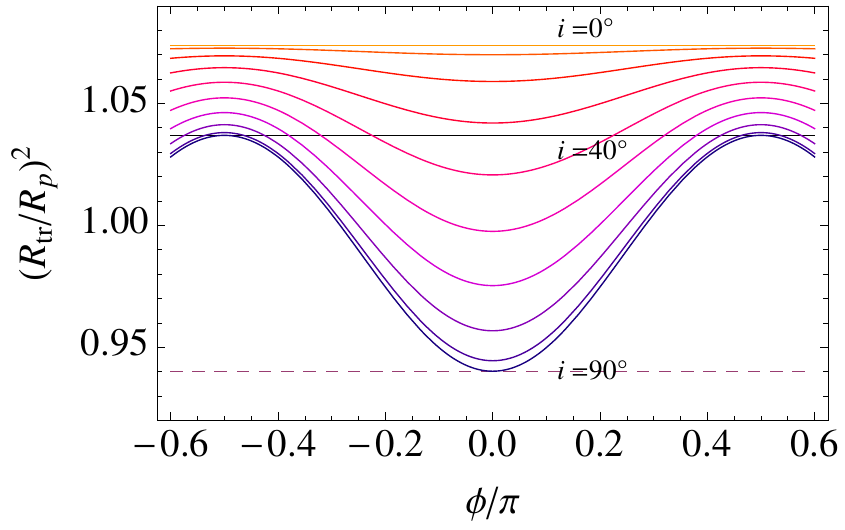} }\\
 \subfigure{ \includegraphics[width=8.5cm]{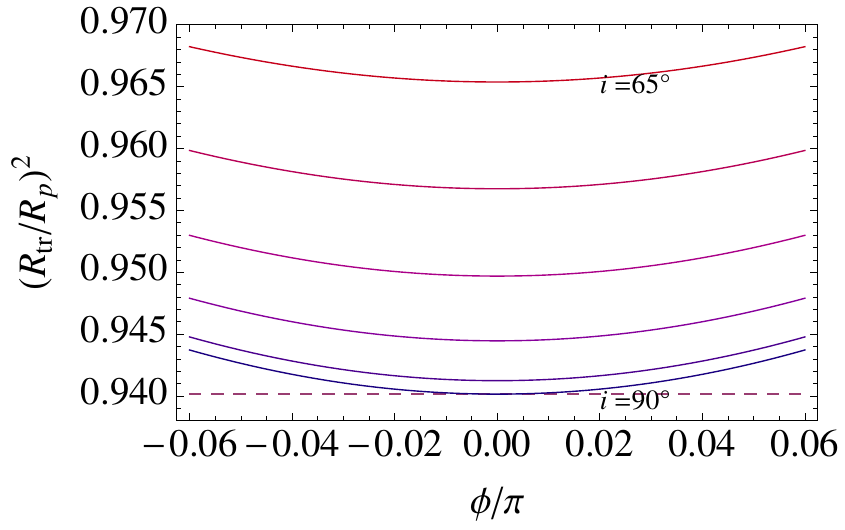}} 
\end{center}
\caption{Normalized projected area of the planet as a function of its anomaly ($\phi$) for inclinations of the orbit going from $i=90^\circ$ (lowest curve) to $i=0^\circ$ (highest curve) by steps of 10$^\circ$ (5$^\circ$ for bottom panel) for a WASP-12\,b analog on a circular orbit. Top: For the full orbit. Bottom: zoom on the (primary or secondary) transit. The ordinates of the dotted, solid and dashed horizontal lines are respectively $a_1a_2/R_0^2$ (face-on orbit), $a_1a_3/R_0^2$ and $a_2a_3/R_0^2$.}
\label{fig:projec}
\end{figure}

For the simple case of an edge-on orbit at mid transit ($\phi=0$, $i=90^\circ$), since the observer, the planet and the star are aligned with the long axis of the tidally deformed ellipsoid\footnote{This is still verified to first order in $\phi$ and $i-\frac{\pi}{2}$ as only second order terms appear.}$^,$\footnote{In the following, the variables have the same meaning as in \sect{sec:variational} and \ref{sec:shape} with p indices when referring to the planet and $\star$ to the star}, $\rtrp=\sqrt{a_{2,\p}a_{3,\p}}$ and $\rtrs=\sqrt{a_{2,\star}a_ {3,\star}}$.
Therefore, 
\begin{align}\label{transitdepth2}
\delta= \frac{a_{2,\p}a_{3,\p}}{a_{2,\star}a_ {3,\star}}=\left(\frac{R_\mathrm{0,p}}{R_{0,\star}}\right)^2\cdot (1+\eta),
\end{align}
where $R_\mathrm{0,p}$ and $R_{0,\star}$ are the respective radii the planet and the star would have in spherical equilibrium and {\it $\eta$ is by definition the variation of the transit depth induced by the ellipsoidal shape of the components relative to the transit depth in the spherical case}. To first order in the deformation, this is given by
\begin{align}
\eta=\alpha_{2,\p}+\alpha_{3,\p}-\alpha_{2,\star}-\alpha_ {3,\star}.
\end{align}
The choice of the expression to be taken for the $\ai$ depends on the physical context. In the general case, one can use a linear combination of \eqs{rotation}{tides} and get a general expression which depends on $r$, $\omega_\p$ and $\omega_\star$. However, most of the planet hosting stars have a low rotation rate compared to the orbital mean motion. This entails that the rotational deformation is negligible compared to the tidal one and can generally be neglected.
As mentioned in \sect{sec:sp}, hot Jupiters should be pseudo synchronized early in their evolution. Therefore, we will assume such an approximation in our calculations in order not
to introduce any other free parameter. The impact of the rotation alone is described \sect{sec:rd}. Under such an approximation,
\begin{align}\label{eta}
\eta=&-\frac{1}{3}\,q_\p\left(\frac{1+p}{ p}\right) \frac{R_{0,\p}^3}{ r^3}\left[\frac{5}{4}
          \left(\frac{7+p}{1+p}\right)-\left(\frac{4\np}{3-\np}\right)\right] \nonumber \\
&+\frac{5}{2}\, q_\star\, p \frac{R_{0,\star}^3}{ r^3},
\end{align}
where the parameter $p$ now denotes the mass ratio $\mp/\ms$, and
$q_\p$ and $q_\star$ are equal to $q_n$ for $n=\np$ and $n=\ns$,
respectively. The first line in the above equation represents the
contribution of the planet, which is always negative (for reasonable
values of $n$). Our line
of sight follows the long axis of the tidal bulge and we see the
minimal cross section of the ellipsoid.  

The contribution of the star is positive and, in most cases,
negligible compared the planet's contribution
because $$\frac{R_{0,\star}^3}{R_{0,\p}^3}\frac{p^2}{1+p}\ll 1,$$ for
a typical system ($10^{-3}$ for a Jupiter-Sun like system). As a
consequence, the results presented hereafter do not depend on
$q_\star$ as long as realistic values of $\ns\in [1.5,\, 3]$ are
taken.

\begin{figure*}[htbp] 
 \sidecaption
\includegraphics[scale=1.3]{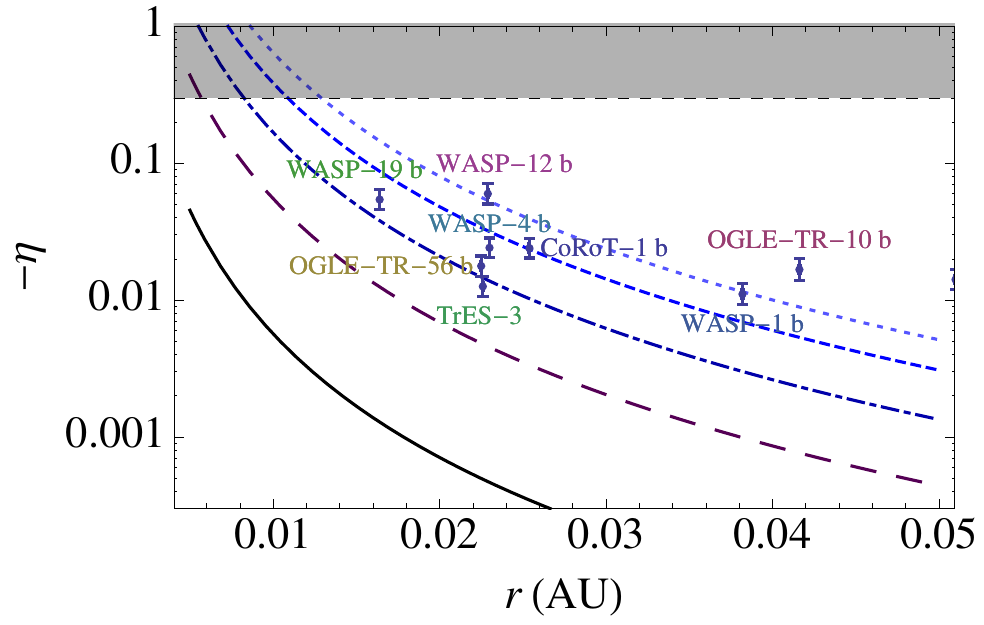}
 \caption{Relative transit depth variation $\eta$ computed with
   \eq{eta} as a function of the semimajor axis at 1 Gyr for planets
   of mass: 0.3$\mjup$ (dotted), 0.5$\mjup$ (dashed), 1$\mjup$
   (dash-dotted), 3$\mjup$ (long dashed), 15$\mjup$(solid). The shaded
   area shows the zone where higher order terms become non-negligible. 
   The decrease of the transit depth due to tidal
   interactions is smaller when the mass of the planet increases
   because massive objects 
%
are more compressible (see \S4) and thus
   less subject to non-spherical deformations.}
 \label{fig:transit1}
\end{figure*}

Figure \ref{fig:transit1} portrays the relative transit depth
variation computed with \eq{eta} for several planet masses as a
function of the orbital distance, for a Sun-like parent star. While
all the curves are calculated at an age of 1 Gyr, they do not change
much for older ages because both the radii and the polytropic indices
remain nearly unchanged after 1 Gyr (see Fig.\,\ref{fig:tr132}). Given
the accuracy of the radius determination achieved by the latest
observations (1 to 10\%), the transit depth variation is significant
for Saturn mass objects ($\mp\approx\mjup/3$) closer than 0.04 AU and
Jupiter mass objects closer than 0.02-0.03 AU. Because we derived the
equations to first order, the value of $\eta$ derived from our model
should be taken with caution when 
$\eta\gtrsim 0.1-0.3$ 
(and are clearly not meaningful for $\eta\gtrsim1$). In this regime,
corresponding to the upper left region of Fig.\,\ref{fig:transit1},
one should use the theory of planetary figures to higher order, but
then numerical calculations become necessary, loosing the advantage of
our simple analytical expressions. Figure \ref{fig:transit1} also
displays the transit depth variation computed for the most distorted
known transiting exoplanets, with the 
observationally measured
parameters. The error bars reflect the uncertainties in the model and
in the measured data.

\subsection{Which radius?}

Before going further, it is important to summarize the differences between the various radii that we have defined above. Note that, in the literature, the term "radius" is used loosely, even for non-spherical objects. Importantly enough, this can lead to discrepant normalizations throughout different studies and published values of transit radius measurements when, for example, radii are shown in units of Jupiter radii ($\rjup$) without precisely defining the latter. 

One can define $a_1$, $a_2$ and $a_3$ as the distances between the center and some isobar surface along the three principal axes of inertia. For any distorted object, we can define the mean radius ($R$) as the radius of the sphere that would enclose the same volume as the described surface. In our case of a general ellipsoid, we have $R=(a_1a_2a_3)^{1/3}$. If axial symmetry holds (e.g. for a rotating fluid body), we have $a_1=a_2\equiv\req$, defining the equatorial radius, and $a_3\equiv\rpol$ the polar radius. Finally, $R_0$ is the radius of the spherical shape that the fluid body would assume if it was isolated and at rest in an inertial frame (the limiting case for which all the mentioned radii would be equal). This latter is the radius computed in usual 1D numerical evolution calculations. Note that in general $R\neq R_0$ because the centrifugal force has a net outward component that increases the volume of the object, as can be seen from \eq{mac1}.

One must be aware that only $a_1$, $a_2$ and $a_3$ (reducing to $\req$ and $\rpol$ for solar system gaseous bodies) can be measured directly and are \textit{not} model dependent. This is why we define $\rjup$ as the equatorial radius of Jupiter at the 1 bar level ($\rjup\equiv R_\mathrm{eq,J}=7.1492\times10^7$m, \citealt{Gui05} and reference therein).

Unfortunately, transit measurements only give access to the projected opaque cross section of the planet ($\equiv\pi \rtrp^2$) defining a "transit radius" which depends on the shape of the planet, its orientation during the observation and the wavelength used. To convert this transit radius inferred from the observations ($\rtrp$) to the spherical radius ($R_{0,\p}$) - that can be compared to 1D numerical models - one must eliminate $\delta$ from \eqs{transitdepth}{transitdepth2}.
As shown above, the stellar impact on $\eta$ is
negligible compared to the planet's contribution ($\rtrs\approx R_{0,\star}$). Then, using the
first term in \eq{eta} and expanding the expression giving the
definition of $\rtrp$, one gets
\begin{align}\label{r0res}
R_{0,\p}&\approx\rtrp\cdot\left(1-\frac{1}{2}(\alpha_{2,\p}+\alpha_{3,\p})\right)\approx\rtrp\cdot\left(1-\frac{\eta}{2}\right).
\end{align}
For the most distorted known planets, the relative variation between
the transit radius and the equilibrium radius $$\Delta R\equiv
(R_{0,\p}-\rtrp)/\rtrp\approx -\eta/2$$ is positive and amounts to
3.00\% for WASP-12\,b, 2.72\% for WASP-19\,b, 1.21\% for WASP-4\,b,
1.20\% for CoRot-1\,b, 0.89\% and OGLE-TR-56\,b.%
\footnote{
Of course, since $
\eta\propto \left(\frac{1+p}{ p}\right) \frac{R_{0,\p}^3}{r^3},
$
 \eq{r0res} is an implicit equation on $R_{0,\p}$. To obtain $R_{0,\p}$ to the sought accuracy, a perturbative development in powers of $\eta_\mathrm{tr}=\eta(R_{0,\p}= \rtrp)$ can be obtained using recursively \eq{r0res}
\begin{align}\label{r0res2}
\frac{R_{0,\p}}{\rtrp}&\approx1-\frac{\eta(R_{0,\p})}{2}\approx1-\frac{\eta_\mathrm{tr}}{2}+\frac{3\,\eta^2_\mathrm{tr}}{4}-\frac{3\,\eta^3_\mathrm{tr}}{2}+\mathcal{O}(\eta^4_\mathrm{tr}).
\end{align}
However, terms of order $\eta^2_\mathrm{tr}$ are of the same order than the second order corrections to the shape that we have neglected throughout.}

Note that because the mean density scales as $R_{0,\p}^{-3}$, the increase in radius implies a decrease in the mean density inferred which is about three times larger (i.e. $\approx 9\%$ for WASP-12\,b). This is of particular importance when one wants to constrain the internal composition or enrichment of giant planets from transit measurements.


\section{Conclusion}\label{sec:conc}

Because of the large variety of exoplanetary systems presently
discovered, with many more expected in the near future, and the
increasing accuracy of the observations, it is important to take into
account the corrections arising from the non-spherical deformation of
the planet or the star, due to rotational and/or tidal forces, as such
a deformation yields a decrease of the transit depth. In order to do
so, it is extremely useful to be able to compute \textit{analytically}
the shape of planets and stars in any configuration from the knowledge
of only their mass, orbital separation and one single parameter
describing their internal structure, namely the polytropic index,
$n$. Such formulae are derived in \sect{sec:shape}, and can be
easily
used to determine the impact of the shape of the planet on
its phase curve and on the shape of the transit light curve itself
\citep{CW10a}. They can also be used to model ellipsoidal variations
of the stellar flux that 
are now detected
in the CoRoT and Kepler light
curves \citep{WOS10}.  These formulae also give good approximations
for various parameters describing the mass redistribution in the
body's interior and the response to a perturbing gravitational field,
i.e. the moment of inertia, $I$, and the Love number of second degree
$k_2$.

Another major implication of the present work is to show that
departure from sphericity of the transiting planets produces a bias in
the determination of the radius. For the closest planets detected so
far ($\lesssim 0.05$ AU), the effect on the transit depth is of the order of
1 to 10\% (see Fig. 4), by no means a negligible effect.  The
equilibrium radius of these strongly distorted objects can thus be
larger than the measured radius, inferred from the area
of the (smaller) cross section presented to the observer by the planet
during the transit. The analytical formulae derived in the present
paper, and the characteristic polytropic index values derived for
various gaseous planet masses and ages, allow to easily take such a
correction into account.  Interestingly, since this equilibrium radius
is the one computed with the 1D structure models available in the
literature, the bias reported here still enhances the magnitude of the
puzzling radius anomaly (see Fig.\,6 of \citet{LCB10}) exhibited by the so-called bloated planets.

\begin{acknowledgements}
The authors are very grateful to Nick Cowan and Louis Shekhtman who first noticed minor errors in Appendix B that have been corrected in the present version.
The authors acknowledge the hospitality of the Kavli Institute for Theoretical
Physics at UCSB (funded by the NSF through Grant PHY05-51164), where
this work started. This work has been supported in part by NASA Grant NNX07AG81G and NSF
grants AST 0707628. We also acknowledge funding from the European Community via the P7/2007-2013 Grant Agreement no. 247060. The authors are grateful to the anonymous referee for his/her sharp and enlightening comments.
\end{acknowledgements}

\bibliography{biblio} 
\bibliographystyle{aa}

\appendix
\section{Theory of planetary figures: \\numerical methods}
\label{theoryFig}

Here, we briefly outline the method described in \cite{Ste39} to compute numerically the response of a body in hydrostatic equilibrium to a perturbing potential\footnote{Here we take the convention that the force acting on a particule of mass $M$ due to a potential $V$ is $\mathbf{F}=-M \mathbf{grad} V$. This yields some difference of signs with \cite{Ste39}.} and derive additional formulae.
To lowest order (which is consistent with the order of approximation used throughout the present paper) the body response is linear and the total deformation is the sum of the response to each term of the decomposition of the perturbing potential. Let us consider a term of the decomposition of the potential of the form
\begin{align}\label{perturbpot}
V_\l^\m(r,\theta,\psi)=c_\l^\m r^\l Y_\l^\m (\theta,\psi),
\end{align}
where the $Y_\l^\m$ are tesseral harmonics defined by 
\begin{align}
Y_\l^\m(\theta,\psi)=\left| \begin{array}{c} \cos  | \m | \psi\\ \sin | \m | \psi \end{array} \right. P_\l^{| \m |}(\cos \theta).
\end{align}
The $\cos$ ($\sin$) corresponds to positive (negative) values of m and $P_\l^\m$ are the usual associated Legendre polynomials. The reference axis defining $\theta$ and $\psi$ may change from one term to the other. For example, the rotation axis is best suited to treat rotational distortion and the line connecting the center of mass of each body is better to describe the tidal distortion. It is shown by \cite{Ste39} that to first order, the shape of the distorted level surface of mean radius $s$ (see \citealt{ZT80} for a detailed definition) takes the form 
\begin{align}
r(s,\theta,\psi)=s \left(1+\s_\l^\m(s)\, Y_\l^\m (\theta,\psi)\right),
\end{align}
where $r$ is the distance between the center and the level surface as a function of $\theta$ and $\psi$ and $\s_\l^\m(s)$ a figure function yet to be calculated.
\cite{Ste39} shown that, ignoring terms of order $\s_\l^\m\times \s_{\l'}^{\m'}$, $\s_\l^\m$ verifies the following differential equation
\begin{align}
\frac{\d^2 \s_\l^\m}{\d s^2}-\frac{\l(\l+1)}{s^2}s_\l^\m+\frac{6}{s^2}\,\frac{ \rho(s)}{\bar{\rho}(s)}\left(s\dd{ \s_\l^\m}{s}+\s_\l^\m\right)=0,
\end{align}
with 
\begin{align}
\bar{\rho}(s)=\frac{3}{s^3}\int_0^s \rho(s') s'^2\d s'.
\end{align}
Using the variable $\eta_\l(s)=\frac{s}{\s_\l^\m(s)}\dd{\s_\l^\m}{s}$, this rewrites
\begin{align}\label{eqdfeta}
s\dd{\eta_\l}{s}+\eta_\l^2-\eta_\l-\l(\l+1) +6\, \frac{\rho(s)}{\bar{\rho}(s)}\left(\eta_\l+1\right)=0.
\end{align}
Then, $\eta_\l(R)$ ($R$ being the external mean radius of the object) can be obtained by numerical integration (with $\eta_\l(0)=\l-2$) and the shape and external potential ($U_\l^\m$) of the body are given by, respectively
\begin{align}
\s_\l^\m(R)=-\frac{2\l+1}{\l+\eta_\l(R)}\frac{1}{M G}c_\l^\m R^{\l+1}
\end{align}
and
\begin{align}\label{extpot}
U_\l^\m(r,\theta,\psi)=\frac{\l+1-\eta_\l(R)}{\l+\eta_\l(R)}\frac{R^{2\l+1}}{r^{\l+1}}c_\l^\m  Y_\l^\m (\theta,\psi).
\end{align}

In order to compare this numerical model with others, we can compute several observable quantities.
By definition, the potential Love number of degree 2 ($k_2$) is given by
\begin{align}\label{LoveNum}
U_2^0(R,\theta,\psi)=k_2 \,V_2^0(R,\theta,\psi),
\end{align}
which yields
\begin{align}
k_2=\frac{3-\eta_2(R)}{2+\eta_2(R)}.
\end{align}
The level Love number ($h_2$) is given by
\begin{align}\label{LoveNum}
\s_2^0(R) \,Y_2^0 (\theta,\psi)=-h_2 \,\frac{V_2^0(R,\theta,\psi)}{g\,R},
\end{align}
where $g$ is the surface gravity acceleration, and
\begin{align}
h_2=\frac{5}{2+\eta_2(R)}=k_2+1,
\end{align}
as expected for a body in hydrostatic equilibrium.

\subsection{Axi-symmetric Case}

Thus the first gravitational moment ($J_2$) defined by
\begin{align}\label{J2def}
U_2^0(r,\theta,\psi)=J_2\frac{G M}{r}\left(\frac{\req}{r}\right)^2P^0_2(\cos \theta),
\end{align}
is given by 
\begin{align}
J_2=k_2\frac{R^3}{G M}c_\l^\m.
\end{align}
No distinction is made between $R$ and $\req$ (the equatorial radius) when comparing \eqs{extpot}{J2def} because this would only add higher order corrections to $J_2$ which is already a first order quantity.

For the rotational distortion of the body whose angular velocity is $\omega$, $c_2^0=\omega^2/3$ and
\begin{align}\label{J2rot}
J_2=\frac{k_2}{3}\frac{\omega^2 R^3}{G M}.
\end{align}
If one is concerned with the external shape, the oblateness ($f$, see \eq{oblate}) of a rotating body is given by
\begin{align}\label{oblateNum}
f&=-\frac{3}{2}\s_2^0 \nonumber\\
&= \frac{h_2}{2}\frac{\omega^2 R^3}{G M}= \frac{k_2+1}{2}\frac{\omega^2 R^3}{G M}.
\end{align}

By extension, one can define $J_2$ for a tidal perturbation by a secondary of mass $M'$ at a distance $r'$, $c_2^0=-\frac{G M'}{r'^3}$ leading to 
\begin{align}\label{J2tides}
J_2=-k_2\frac{ M'}{M}\left(\frac{R}{r'}\right)^3,
\end{align}
but the reference axis is the line connecting the two center of mass and not the rotational axis.

\subsection{Triaxial Case}

While it is tempting to add \eqs{J2rot}{J2tides} to obtain the total $J_2$ of a body in a close binary, we must remember that the tidal and rotational deformations do not have the same axis of symmetry in general. Taking $\theta$ as the colatitude and $\psi$ as the longitude of the body considered,
the external gravitational field of the latter reads
\begin{align}
U(r,&\,\theta,\psi)=-\frac{G M}{r} \times \nonumber \\
&\sum_{\l=0}^\infty\sum_{\m=0}^{\l}\left(\frac{\req}{r}\right)^\l (C_\l^\m\,\cos\m \psi+S_\l^\m\,\sin \m\psi)P^\m_\l(\cos \theta).
\end{align}
 The quadrupole moment in the linear approximation, is given at the surface by
\begin{align}
-\frac{G M}{\req} &\sum_{\m=0}^{2} (C_2^\m\,\cos\m \psi+S_2^\m\,\sin \m\psi)P^\m_2(\cos \theta)\nonumber \\
&=k_2 \req^2\left(\frac{\omega^2}{3}P^0_2(\cos \theta)-\frac{GM'}{r'^3} P^0_2(\cos \theta')\right),
\end{align}
where $\theta'$ is the angle between the current point and the line connecting the two center of mass. For the coplanar case where the tides raising object orbits in the equatorial plane of the distorted body, $\cos \theta' =\sin \theta \cos \psi$ and thus
\begin{align}
P^0_2(\cos \theta')&=\frac{3}{4} \sin ^2(\theta)  \cos (2 \psi )+\frac{3 \sin ^2(\theta )}{4}-\frac{1}{2} \nonumber\\
&=\frac{1}{4}  \cos (2 \psi )P^2_2(\cos \theta)-\frac{1}{2} P^0_2(\cos \theta).
\end{align}
Thus
\begin{align}
J_2=-C_2^0=k_2\left[\frac{1}{3}\frac{\omega^2 \req^3}{G M}+\frac{1}{2}\frac{ M'}{M}\left(\frac{R}{r'}\right)^3\right],
\end{align}
and
\begin{align}
C_2^2=-k_2\frac{1}{4}\frac{ M'}{M}\left(\frac{R}{r'}\right)^3.
\end{align}
All the other moments are equal to 0. Similar decompositions can be used to infer the precise shape of the surface from a sum of perturbing fields. This gives
\begin{align}
\frac{r(R,\theta,\psi)}{R}&=1\nonumber \\
&-h_2 \left(\frac{1}{2}\frac{ M'}{M}\left(\frac{R}{r'}\right)^3+\frac{1}{3}\frac{\omega^2 R^3}{ G M}\right)\,P_2^0 (\cos\theta)\nonumber \\
&+h_2\left(\frac{1}{4}\frac{ M'}{M}\left(\frac{R}{r'}\right)^3\right)\,\cos (2\psi)\,P_2^2 (\cos\theta),
\end{align}
which directly translates into $a_1$, $a_2$ and $a_3$ (once $R$ is known) by setting $(\theta,\psi)$ equal to $(\pi/2,0)$, $(\pi/2,\pi/2)$ and $(0,0)$, respectively. Translating this into $\alpha_1$, $\alpha_2$ and $\alpha_3$ is a little more complicated because one needs to account for the fact that $R>R_0$ due to the centrifugal potential. This can be taken into account either numerically - by including the centrifugal force when solving the hydrostatic equilibrium - or analytically using \eq{rotexpansion} or (\ref{rotexpansion2}).

\section{Projected area of a triaxial ellipsoid}
\label{geometry}

\subsection{General case}

Let us define two coordinate systems. The first one ($\mathbf{\hat{x}'}$, $\mathbf{\hat{y}'}$, $\mathbf{\hat{z}'}$) is defined by the three main axes of the ellipsoid.
In this frame, the equation of the surface of the ellipsoid is
 \begin{align}
\frac{x'^2}{a^2_1}+\frac{y'^2}{a^2_2}+\frac{z'^2}{a^2_3}=1.
\end{align}
To compute the projected area of this ellipsoid as it will be seen by the observer, it is easier to put ourselves in another coordinate system defined by 
the line connecting the center of mass of the system and the observer (toward the observer; $\mathbf{\hat{x}}$), the projection of the orbital angular momentum on the sky plane ($\mathbf{\hat{z}}$) and a third axis in the sky plane chosen so that $(\mathbf{\hat{x}},\,\mathbf{\hat{y}},\,\mathbf{\hat{z}})$ follows the right-hand vector sense. The current position vector $(\vr=(x,\,y,\,z))$ expressed in this frame is thus related to the one expressed in the first coordinate system by a rotation matrix $ \mathcal{R}$ such as
 \begin{align}\label{Rmatrix}
 \mathbf{r'}= \mathcal{R}\cdot\vr,
\end{align}
With $\mathcal{R}^\mathrm{tr}\mathcal{R}=\mathbf{1}$.
The equation of the ellipsoid in the new system thus writes 
 \begin{align}\label{genellips}
g(\vr)\equiv\vr^\mathrm{tr}\,\mathcal{R}^\mathrm{tr}\,\left( \begin{array}{ccc} \frac{1}{a_1^2} & 0 & 0 \\ 0 &  \frac{1}{a_2^2} & 0 \\ 0 & 0 & \frac{1}{a_3^2} \end{array}\right) \,\mathcal{R}\,\vr\equiv \vr^\mathrm{tr}\,\mathcal{A}\,\vr=1.
\end{align}
The exact value of the matrix $\mathcal{A}$ will depend on the rotation needed and on the angles chosen to represent it. This can be worked out in each specific case. To keep some generality, we will take $\mathcal{A}$ of the form
 \begin{align}
\mathcal{A}=\left( \begin{array}{ccc} a & d & f \\ d & b & e \\ f & e & c \end{array}\right).
\end{align}
The symmetry is ensured by the fact that both of our coordinate systems are orthonormal.
The equation of the contour of the projected shadow is given by the fact that the normal to the ellipsoid is normal to the line of sight ($\mathbf{\hat{x}}$) there. This assumes a completely opaque body below the isobar chosen to be the surface. The complete calculation of the level at which optical rays that are grazing, at the terminator, have an optical depth close to unity \citep{HFL01,BSH03,Gui10} - in the present geometry - should give rise to subtle effects but of smaller importance. This reads
 \begin{align}
0&=\mathbf{grad} [g(\vr)]^\mathrm{tr} \cdot \mathbf{\hat{x}}\nonumber \\
&=2 \,\vr^\mathrm{tr}\, \mathcal{A}\,\mathbf{\hat{x}}.
\end{align}
This shows that these points are located on a plane whose equation is (since $a\neq0$)
 \begin{align}\label{plane}
x=-\frac{1}{a}(d\,y+f\,z).
\end{align}
Substituting $x$ in \eq{genellips} by \eq{plane} we see that the cross section is an ellipse following the equation
 \begin{align}
(y,z)\,\left( \begin{array}{cc}  b-\frac{d^2}{a} & e-\frac{df}{a} \\  e-\frac{df}{a} & c-\frac{f^2}{a} \end{array}\right)\,\left( \begin{array}{c} y\\ z \end{array}\right)\equiv (y,z)\,\mathcal{B}\,\left( \begin{array}{c} y\\ z \end{array}\right)=1.
\end{align}
It is thus possible to find the rotation in the sky plane needed to reduce the ellipse and find its principal axes ($p_1,\,p_2$). If only the cross section ($\pi p_1 p_2$) is needed, we can use the fact that the determinant of a matrix is independent of the coordinate system so that
\begin{align}
\pi p_1 p_2=\frac{\pi}{\sqrt{\mathrm{Det}(\mathcal{B})}},
\end{align}
with
\begin{align}
\mathrm{Det}(\mathcal{B})=b c-e^2-\frac{ b f^2}{a}-\frac{ c d^2}{a}+2\,\frac{ d e f}{a}.
\end{align}
In the case of an edge-on orbit at mid transit, no rotation is needed, $\mathcal{R}$ is the identity and thus $a=1/a^2_1$, $b=1/a^2_2$, $c=1/a^2_3$ and $d=e=f=0$. We retrieve 
\begin{align}
\pi p_1 p_2=\frac{\pi}{\sqrt{bc}}=\pi a_2 a_3.
\end{align}

\subsection{Coplanar case}

If the planet equator and the orbital plane are coplanar, the unit vectors of first coordinate system defined above coincides with the unit vectors defined by the line connecting the two center of mass (from the secondary to the object under consideration; $\mathbf{\hat{x}'}$), its normal in the orbital plane (in the direction of motion; $\mathbf{\hat{y}'}$) and the rotation axis of the body ($\mathbf{\hat{z}'}$). If $i$ is the inclination of the orbit with respect to the sky plane and $\phi$ the true anomaly defined to be 0 at mid transit, the rotation matrix defined by \eq{Rmatrix} reads
\begin{align}
\mathcal{R}=\left(
\begin{array}{ccc}
 \sin (i) \cos (\phi ) & \sin (\phi ) & \cos (i) \cos (\phi ) \\
 -\sin (i) \sin (\phi ) & \cos (\phi ) & -\cos (i) \sin (\phi ) \\
 -\cos (i) & 0 & \sin (i)
\end{array}
\right).
\end{align}

The $\mathcal{A}$ matrix can be computed thanks to \eq{genellips} giving the $a$, $b$, ..., $f$ coefficients and thus $\mathrm{Det}(\mathcal{B})$. This gives the project area of the planet or the star at any given point of the orbit
\begin{align}
\pi p_1 p_2=\pi  \sqrt{a_3^2 \sin ^2i \left(a_1^2 \sin ^2\phi+a_2^2 \cos ^2\phi\right)+a_1^2
   a_2^2 \cos ^2i}.
\end{align}
as shown on Fig.\,\ref{fig:projec}. Note however, that this formula is more general. It is also the cross section that would be seen by an observer located in the direction
\begin{align}\vr_\mathrm{obs}=\left( \begin{array}{ccc}\sin i \cos \phi\\\sin i \sin \phi\\\cos i\end{array}\right) ,\end{align}
in the reference frame defined by the three main axes of the ellipsoid $(\mathbf{\hat{x}},\,\mathbf{\hat{y}},\,\mathbf{\hat{z}})$.

\begin{table*}[p]
\begin{center}
\caption{
Summary of variables}
\label{tab:cond}
\small
\begin{tabular}{ c l l} \hline\hline 
Symbol & Definition & Equation \\ \hline
 \hline
$E$ & total energy & \ref{nrj} \\
$U$ & internal energy & \ref{Uint}\\
$W$ & self gravitational energy & \ref{Wself}\\
$T$ & total kinetic energy & \ref{totalT}\\
$T_\mathrm{s}$ & rotational kinetic energy & \ref{Ts}\\
$T_0$ & orbital kinetic energy & \ref{T0}\\
$W_\i$ & gravitational interaction energy & \ref{interpotential}\\
$n,~K$ & polytropic index and temperature & \ref{poly}\\
$G$& gravitational constant &\\
$M$& Mass of primary &\\
$R$& Mean radius of the primary &\\
$P$& Pressure &\\
$\rho$ & density &\\
$a_\i$ & principal axes of the ellipsoid & \\
$\alpha_\i$ & first order correction to $a_\i$ & \ref{defalpha}\\
$r$ & orbital distance &\\
$\lambda_\i$& asymmetry factor $=(a_3/a_\i)^{2/3}$&\\
$\rhoc$& central density & \ref{Uint}\\
$\bar{k}_1$& scaling constant of the internal energy & \ref{Uint}, \ref{k1}\\
$\bar{k}_2$& scaling constant of the gravitational energy & \ref{Wself}, \ref{k2}\\
$\bar{f}$& ellipsoidal correction to the gravitational energy & \ref{Wself}, \ref{fbar}\\
$A_\i$ && \ref{Ai}\\
$I$& principal moment of inertia &\ref{defI}, \ref{dimensionlessmoment}\\
$\kappa_n$& Dimensionless moment of inertia & \ref{defkappa}\\
$\Omega$& orbital mean motion & \ref{T0}, \ref{kepler}\\
$\Lambda$& internal angular velocity of the fluid & \ref{Ts}\\
$\omega$ & rotational angular velocity & \ref{omega}\\
$\xi_1$& dimensionless radius (See \citealt{Ch39})& \ref{k1}\\
$\theta'_1$& dimensionless density derivative (See \citealt{Ch39})& \ref{k1}\\
$I_{\i\j}$& Inertia tensor &\ref{Itensor}\\
$\delta_{\i\j}$& Kronecker Symbol &\\
$\Delta$&& \ref{kepler}\\
$g_\mathrm{t}$&& \ref{g_t}\\
$R_0$& radius of the \textit{unperturbed} spherical polytrope & \ref{spherical_Radius}\\
$q_n$ &$=\kappa_n(1-\frac{n}{5})$ & \ref{shape1}, \ref{shape2}\\
$\mur$ & $=GM'/r^3$& \ref{shape1}, \ref{shape2}\\
$\murt$& $\mur/(\pi G \bar{\rho})$&\ref{shape1}, \ref{shape2}\\
$\bar{\rho}$& mean density &\\
$\bar{\omega}$& dimensionless angular velocity & \ref{omegabar} \\
$f$& oblateness & \ref{oblate} \\
$J_2$& quadrupolar gravitational moment & \ref{J2}\\
$k_2$&  Love number (half the apsidal motion constant) & \ref{lovenumber}\\
$ L_\star$& stellar luminosity out of transit & \ref{transitdepth}\\
$\delta L_\star$& stellar luminosity variation in transit & \ref{transitdepth}\\
$\delta$&relative transit depth & \ref{transitdepth}\\
$R_\mathrm{tr}$& transit radius & \ref{transitdepth} \\
$\eta$& non spherical contribution to the transit depth & \ref{transitdepth2}\\
$$&&\\
$$&&\\
\hline\hline
\end{tabular}
\normalsize
\end{center}
\end{table*}
\clearpage

\clearpage

\begin{table*}[p]
\begin{center}
\caption{
Model parameters for non-irradiated planets of various masses. 
For each planetary mass ($\mp$) and age, this table gives the
spherical equilibrium radius ($R_\mathrm{0,p}$, $\rjup\equiv 7.1492\times10^7$m), the effective
temperature ($\ind{T}{eff}$), the central pressure ($\ind{P}{c}$),
temperature ($\ind{T}{c}$) and density ($\ind{\rho}{c}$) along with
the polytropic index ($n$), the dimensionless moment of inertia
($\kappa_n$, see \eq{dimensionlessmoment}) and Love number ($k_2$). $k_2$ was derived both with our analytical formula (\eq{lovenumber}; analytic) and with a numerical method (See Appendix \ref{theoryFig}; numeric). An electronic version of this table is available at \textit{http://perso.ens-lyon.fr/jeremy.leconte/JLSite/JLsite /Exoplanets\_Simulations.html}}
\label{tab:cond}
\small
\begin{tabular}{ r@{.}l r@{.}l r@{.}l r@{.}l r@{.}l r@{.}l r@{.}l  r@{.}l r@{.}l  r@{.}l | r@{.}l} \hline\hline 
\multicolumn{2}{c}{$\mp\,(\mjup)$}	&\multicolumn{2}{c}{Age (Gyr)} &\multicolumn{2}{c}{$R_\mathrm{0,p}\,(\rjup)$}  & \multicolumn{2}{c}{$\ind{T}{eff}\,(K)$} & \multicolumn{2}{c}{$\ind{P}{c}\,$(M\,Bar)} &\multicolumn{2}{c}{$\ind{T}{c}\,(K)$ }& \multicolumn{2}{c}{$\ind{\rho}{c}\,$(kg/m$^3$)} & \multicolumn{2}{c}{$n$} & \multicolumn{2}{c}{$\kappa_n$} & \multicolumn{4}{c}{$k_2$} \\
 \multicolumn{18}{c}{ } & \multicolumn{2}{c}{analytic}& \multicolumn{2}{c}{numeric} \\ \hline
 \hline
     0&35 &      0&05 &    1&1579 &       235&6 &           2&87 &       15\,324& &          1\,016& &        0&913 &    0&681 &    0&568 &    0&428\\
     0&35 &      0&10 &    1&0923 &       194&6 &           3&26 &       13\,566& &          1\,109& &        0&868 &    0&694 &    0&598 &    0&478\\
     0&35 &      0&50 &    0&9934 &       118&4 &           4&05 &        9\,698& &          1\,292& &        0&801 &    0&715 &    0&645 &    0&578\\
     0&35 &      1&00 &    0&9801 &        93&4 &           4&18 &        9\,142& &          1\,326& &        0&792 &    0&718 &    0&651 &    0&594\\
     0&35 &      5&00 &    0&9501 &        58&9 &           4&50 &        7\,758& &          1\,402& &        0&771 &    0&725 &    0&667 &    0&631\\
     0&5 &      0&05 &    1&1730 &       280&9 &           5&40 &       19\,045& &          1\,394& &        0&946 &    0&670 &    0&547 &    0&445\\
     0&5 &      0&10 &    1&1112 &       234&8 &           6&06 &       16\,765& &          1\,507& &        0&897 &    0&686 &    0&579 &    0&495\\
     0&5 &      0&50 &    1&0165 &       140&2 &           7&47 &       12\,120& &          1\,742& &        0&838 &    0&704 &    0&618 &    0&578\\
     0&5 &      1&00 &    0&9961 &       108&4 &           7&91 &       10\,950& &          1\,811& &        0&834 &    0&705 &    0&621 &    0&595\\
     0&5 &      5&00 &    0&9682 &        62&4 &           8&53 &        9\,160& &          1\,918& &        0&824 &    0&708 &    0&628 &    0&624\\
     0&6 &      0&05 &    1&1809 &       302&0 &           7&67 &       21\,971& &          1\,670& &        0&987 &    0&658 &    0&522 &    0&443\\
     0&6 &      0&10 &    1&1233 &       254&0 &           8&51 &       19\,377& &          1\,795& &        0&936 &    0&674 &    0&553 &    0&490\\
     0&6 &      0&50 &    1&0292 &       153&9 &          10&44 &       13\,816& &          2\,067& &        0&869 &    0&694 &    0&597 &    0&571\\
     0&6 &      1&00 &    1&0063 &       120&0 &          11&06 &       12\,332& &          2\,146& &        0&861 &    0&697 &    0&602 &    0&591\\
     0&6 &      5&00 &    0&9775 &        66&7 &          12&01 &       10\,196& &          2\,278& &        0&856 &    0&698 &    0&606 &    0&616\\
     0&7 &      0&05 &    1&1842 &       322&0 &          10&44 &       25\,065& &          1\,945& &        1&021 &    0&648 &    0&501 &    0&440\\
     0&7 &      0&10 &    1&1313 &       271&9 &          11&50 &       22\,040& &          2\,090& &        0&974 &    0&662 &    0&529 &    0&484\\
     0&7 &      0&50 &    1&0390 &       166&9 &          14&00 &       15\,781& &          2\,398& &        0&901 &    0&684 &    0&576 &    0&562\\
     0&7 &      1&00 &    1&0155 &       129&9 &          14&79 &       13\,766& &          2\,492& &        0&887 &    0&688 &    0&585 &    0&583\\
     0&7 &      5&00 &    0&9849 &        71&0 &          16&13 &       11\,251& &          2\,644& &        0&882 &    0&690 &    0&589 &    0&607\\
     0&8 &      0&05 &    1&1857 &       339&9 &          13&64 &       28\,043& &          2\,218& &        1&044 &    0&641 &    0&488 &    0&439\\
     0&8 &      0&10 &    1&1360 &       285&7 &          14&98 &       24\,692& &          2\,374& &        1&003 &    0&654 &    0&512 &    0&478\\
     0&8 &      0&50 &    1&0466 &       178&8 &          18&13 &       17\,626& &          2\,728& &        0&929 &    0&676 &    0&558 &    0&553\\
     0&8 &      1&00 &    1&0224 &       139&5 &          19&22 &       15\,272& &          2\,846& &        0&912 &    0&681 &    0&568 &    0&574\\
     0&8 &      5&00 &    0&9903 &        76&1 &          20&96 &       12\,243& &          3\,022& &        0&903 &    0&684 &    0&575 &    0&599\\
     0&9 &      0&05 &    1&1864 &       355&5 &          17&22 &       30\,999& &          2\,469& &        1&057 &    0&637 &    0&480 &    0&438\\
     0&9 &      0&10 &    1&1393 &       297&1 &          18&84 &       27\,295& &          2\,641& &        1&021 &    0&648 &    0&501 &    0&475\\
     0&9 &      0&50 &    1&0525 &       189&6 &          22&83 &       19\,399& &          3\,048& &        0&952 &    0&669 &    0&543 &    0&546\\
     0&9 &      1&00 &    1&0272 &       148&9 &          24&30 &       16\,774& &          3\,191& &        0&933 &    0&674 &    0&555 &    0&567\\
     0&9 &      5&00 &    0&9946 &        82&0 &          26&53 &       13\,201& &          3\,402& &        0&921 &    0&678 &    0&563 &    0&592\\
     1&0 &      0&05 &    1&1873 &       369&1 &          21&02 &       33\,873& &          2\,696& &        1&061 &    0&636 &    0&478 &    0&440\\
     1&0 &      0&10 &    1&1419 &       306&5 &          23&04 &       29\,826& &          2\,897& &        1&030 &    0&645 &    0&496 &    0&474\\
     1&0 &      0&50 &    1&0576 &       199&5 &          27&98 &       21\,165& &          3\,357& &        0&970 &    0&663 &    0&532 &    0&540\\
     1&0 &      1&00 &    1&0316 &       157&6 &          29&88 &       18\,218& &          3\,528& &        0&951 &    0&669 &    0&544 &    0&560\\
     1&0 &      5&00 &    0&9980 &        87&0 &          32&77 &       14\,132& &          3\,779& &        0&936 &    0&673 &    0&553 &    0&585\\
     1&2 &      0&05 &    1&1902 &       397&7 &          29&16 &       39\,189& &          3\,127& &        1&051 &    0&639 &    0&484 &    0&445\\
     1&2 &      0&10 &    1&1467 &       328&1 &          32&22 &       34\,685& &          3\,384& &        1&032 &    0&645 &    0&495 &    0&475\\
     1&2 &      0&50 &    1&0666 &       216&9 &          39&33 &       24\,692& &          3\,949& &        0&992 &    0&657 &    0&519 &    0&530\\
     1&2 &      1&00 &    1&0396 &       173&3 &          42&28 &       21\,047& &          4\,173& &        0&975 &    0&662 &    0&529 &    0&550\\
     1&2 &      5&00 &    1&0037 &        96&0 &          46&87 &       15\,917& &          4\,513& &        0&960 &    0&666 &    0&538 &    0&574\\
     1&5 &      0&05 &    1&1974 &       437&2 &          43&27 &       46\,493& &          3\,761& &        1&019 &    0&648 &    0&502 &    0&456\\
     1&5 &      0&10 &    1&1541 &       362&0 &          48&34 &       41\,254& &          4\,096& &        1&016 &    0&650 &    0&504 &    0&479\\
     1&5 &      0&50 &    1&0768 &       238&7 &          59&61 &       29\,797& &          4\,811& &        0&998 &    0&655 &    0&515 &    0&524\\
     1&5 &      1&00 &    1&0496 &       193&9 &          64&49 &       25\,342& &          5\,108& &        0&989 &    0&658 &    0&520 &    0&541\\
     1&5 &      5&00 &    1&0103 &       108&2 &          72&65 &       18\,523& &          5\,584& &        0&976 &    0&661 &    0&528 &    0&563\\
     1&8 &      0&05 &    1&2044 &       467&7 &          59&79 &       53\,447& &          4\,405& &        0&994 &    0&656 &    0&517 &    0&465\\
     1&8 &      0&10 &    1&1613 &       392&9 &          67&23 &       47\,612& &          4\,816& &        0&999 &    0&655 &    0&514 &    0&484\\
     1&8 &      0&50 &    1&0853 &       257&2 &          83&49 &       34\,881& &          5\,679& &        0&997 &    0&655 &    0&515 &    0&521\\
     1&8 &      1&00 &    1&0575 &       211&6 &          90&76 &       29\,639& &          6\,049& &        0&993 &    0&656 &    0&518 &    0&535\\
     1&8 &      5&00 &    1&0157 &       119&2 &         103&42 &       21\,166& &          6\,665& &        0&984 &    0&659 &    0&524 &    0&556\\
     2&1 &      0&05 &    1&2098 &       493&7 &          79&29 &       60\,248& &          5\,071& &        0&980 &    0&660 &    0&526 &    0&471\\
     2&1 &      0&10 &    1&1680 &       420&3 &          89&18 &       54\,147& &          5\,542& &        0&989 &    0&658 &    0&521 &    0&487\\
     2&1 &      0&50 &    1&0920 &       272&9 &         111&54 &       39\,939& &          6\,568& &        0&996 &    0&656 &    0&516 &    0&518\\
     2&1 &      1&00 &    1&0639 &       227&4 &         121&62 &       33\,971& &          7\,009& &        0&995 &    0&656 &    0&517 &    0&530\\
     2&1 &      5&00 &    1&0199 &       129&6 &         139&84 &       23\,851& &          7\,773& &        0&988 &    0&658 &    0&521 &    0&550\\
     3&0 &      0&05 &    1&2203 &       578&7 &         156&52 &       83\,329& &          7\,108& &        0&980 &    0&660 &    0&526 &    0&480\\
     3&0 &      0&10 &    1&1803 &       488&6 &         176&70 &       74\,971& &          7\,806& &        0&989 &    0&658 &    0&520 &    0&490\\
     3&0 &      0&50 &    1&1031 &       316&3 &         225&68 &       55\,401& &          9\,414& &        1&006 &    0&652 &    0&510 &    0&510\\
     3&0 &      1&00 &    1&0756 &       266&8 &         246&76 &       47\,633& &         10\,073& &        1&009 &    0&652 &    0&508 &    0&518\\
     3&0 &      5&00 &    1&0264 &       158&3 &         290&80 &       32\,425& &         11\,396& &        1&009 &    0&652 &    0&509 &    0&534\\
\hline\hline
\end{tabular}
\normalsize
\end{center}
\end{table*}
\clearpage

\begin{table}[p]
\begin{center}
\caption{End of Table\,\ref{tab:cond}.}
\label{tab:cond2}
\small
\begin{tabular}{ r@{.}l r@{.}l r@{.}l r@{.}l r@{.}l r@{.}l r@{.}l  r@{.}l r@{.}l  r@{.}l | r@{.}l} \hline\hline 
\multicolumn{2}{c}{$\mp\,(\mjup)$}	&\multicolumn{2}{c}{Age (Gyr)} &\multicolumn{2}{c}{$R_\mathrm{0,p}\,(\rjup)$}  & \multicolumn{2}{c}{$\ind{T}{eff}\,(K)$} & \multicolumn{2}{c}{$\ind{P}{c}\,$(M\,Bar)} &\multicolumn{2}{c}{$\ind{T}{c}\,(K)$ }& \multicolumn{2}{c}{$\ind{\rho}{c}\,$(kg/m$^3$)} & \multicolumn{2}{c}{$n$} & \multicolumn{2}{c}{$\kappa_n$} & \multicolumn{4}{c}{$k_2$} \\
 \multicolumn{18}{c}{ } & \multicolumn{2}{c}{analytic}& \multicolumn{2}{c}{numeric} \\ \hline \hline
     5&0 &      0&05 &    1&2306 &       746&7 &         437&22 &      134\,158& &         12\,207& &        1&038 &    0&643 &    0&491 &    0&477\\
     5&0 &      0&10 &    1&1855 &       620&1 &         509&10 &      121\,231& &         13\,722& &        1&053 &    0&638 &    0&483 &    0&479\\
     5&0 &      0&50 &    1&1058 &       401&2 &         671&34 &       92\,480& &         16\,920& &        1&076 &    0&631 &    0&469 &    0&485\\
     5&0 &      1&00 &    1&0775 &       333&8 &         741&09 &       80\,859& &         18\,218& &        1&081 &    0&630 &    0&467 &    0&489\\
     5&0 &      5&00 &    1&0226 &       214&3 &         900&43 &       54\,736& &         21\,064& &        1&085 &    0&629 &    0&464 &    0&498\\
     8&0 &      0&05 &    1&2338 &       986&7 &      1\,202&57 &      222\,136& &         20\,971& &        1&140 &    0&612 &    0&434 &    0&453\\
     8&0 &      0&10 &    1&1759 &       804&3 &      1\,468&06 &      201\,343& &         24\,340& &        1&159 &    0&607 &    0&425 &    0&452\\
     8&0 &      0&50 &    1&0854 &       516&0 &      2\,028&59 &      154\,227& &         30\,946& &        1&178 &    0&601 &    0&415 &    0&452\\
     8&0 &      1&00 &    1&0568 &       428&1 &      2\,248&11 &      135\,487& &         33\,390& &        1&179 &    0&601 &    0&414 &    0&454\\
     8&0 &      5&00 &    1&0018 &       278&2 &      2\,740&70 &       94\,044& &         38\,622& &        1&175 &    0&602 &    0&416 &    0&461\\
    10&0 &      0&05 &    1&2355 &    1\,131&0 &      1\,940&74 &      286\,246& &         27\,042& &        1&191 &    0&598 &    0&408 &    0&439\\
    10&0 &      0&10 &    1&1682 &       924&2 &      2\,439&07 &      259\,634& &         32\,036& &        1&206 &    0&593 &    0&401 &    0&438\\
    10&0 &      0&50 &    1&0703 &       582&7 &      3\,458&94 &      198\,170& &         41\,464& &        1&217 &    0&590 &    0&395 &    0&437\\
    10&0 &      1&00 &    1&0415 &       482&5 &      3\,839&83 &      174\,150& &         44\,787& &        1&215 &    0&591 &    0&396 &    0&439\\
    10&0 &      5&00 &    0&9873 &       313&6 &      4\,676&54 &      121\,844& &         51\,786& &        1&206 &    0&593 &    0&400 &    0&445\\
    13&0 &      0&05 &    1&2413 &    1\,330&1 &      3\,355&37 &      388\,856& &         35\,933& &        1&243 &    0&583 &    0&383 &    0&422\\
    13&0 &      0&10 &    1&1605 &    1\,099&6 &      4\,390&75 &      352\,475& &         43\,831& &        1&249 &    0&581 &    0&379 &    0&421\\
    13&0 &      0&50 &    1&0501 &       684&2 &      6\,502&34 &      268\,237& &         58\,583& &        1&247 &    0&581 &    0&380 &    0&422\\
    13&0 &      1&00 &    1&0203 &       559&1 &      7\,258&53 &      234\,464& &         63\,563& &        1&243 &    0&583 &    0&383 &    0&423\\
    13&0 &      5&00 &    0&9677 &       361&2 &      8\,837&13 &      166\,014& &         73\,551& &        1&231 &    0&586 &    0&388 &    0&429\\
    17&0 &      0&05 &    1&5180 &    1\,963&5 &      2\,728&85 &      566\,417& &         27\,110& &        1&258 &    0&578 &    0&375 &    0&403\\
    17&0 &      0&10 &    1&2104 &    1\,460&9 &      6\,602&33 &      518\,481& &         52\,312& &        1&287 &    0&570 &    0&362 &    0&405\\
    17&0 &      0&50 &    1&0308 &       829&1 &     12\,389&70 &      373\,005& &         83\,468& &        1&270 &    0&575 &    0&370 &    0&408\\
    17&0 &      1&00 &    0&9989 &       672&8 &     13\,988&06 &      325\,053& &         91\,373& &        1&264 &    0&577 &    0&373 &    0&410\\
    17&0 &      5&00 &    0&9441 &       421&3 &     17\,311&92 &      230\,449& &        107\,176& &        1&253 &    0&580 &    0&378 &    0&415\\
    20&0 &      0&05 &    1&3776 &    1\,899&6 &      5\,422&33 &      663\,050& &         42\,143& &        1&301 &    0&566 &    0&356 &    0&396\\
    20&0 &      0&10 &    1&1891 &    1\,528&5 &      9\,683&22 &      607\,482& &         64\,873& &        1&303 &    0&566 &    0&355 &    0&397\\
    20&0 &      0&50 &    1&0200 &       903&9 &     17\,617&27 &      446\,001& &        101\,492& &        1&281 &    0&572 &    0&365 &    0&401\\
    20&0 &      1&00 &    0&9827 &       731&4 &     20\,350&09 &      387\,102& &        113\,070& &        1&275 &    0&574 &    0&368 &    0&403\\
    20&0 &      5&00 &    0&9291 &       458&3 &     25\,212&93 &      276\,207& &        132\,824& &        1&265 &    0&576 &    0&372 &    0&407\\\hline\hline
\end{tabular}
\normalsize
\end{center}
\end{table}

\clearpage

\begin{table}[p]
\begin{center}
\caption{End of Table\,\ref{tab:cond}.}
\label{tab:cond3}
\small
\begin{tabular}{ r@{.}l r@{.}l r@{.}l r@{.}l r@{.}l r@{.}l r@{.}l  r@{.}l r@{.}l  r@{.}l | r@{.}l} \hline\hline 
\multicolumn{2}{c}{$\mp\,(\mjup)$}	&\multicolumn{2}{c}{Age (Gyr)} &\multicolumn{2}{c}{$R_\mathrm{0,p}\,(\rjup)$}  & \multicolumn{2}{c}{$\ind{T}{eff}\,(K)$} & \multicolumn{2}{c}{$\ind{P}{c}\,$(M\,Bar)} &\multicolumn{2}{c}{$\ind{T}{c}\,(K)$ }& \multicolumn{2}{c}{$\ind{\rho}{c}\,$(kg/m$^3$)} & \multicolumn{2}{c}{$n$} & \multicolumn{2}{c}{$\kappa_n$} & \multicolumn{4}{c}{$k_2$} \\
 \multicolumn{18}{c}{ } & \multicolumn{2}{c}{analytic}& \multicolumn{2}{c}{numeric} \\ \hline
 \hline
     25& &      0&05 &    1&5200 &     2\,127& &        5\,955& &      815\,455& &         40\,447& &        1&287 &    0&570 &    0&362 &    0&395\\
     25& &      0&10 &    1&2578 &     1\,787& &       12\,709& &      795\,339& &         71\,202& &        1&304 &    0&565 &    0&354 &    0&396\\
     25& &      0&50 &    1&0257 &     1\,044& &       28\,841& &      590\,541& &        131\,050& &        1&305 &    0&565 &    0&354 &    0&389\\
     25& &      1&00 &    0&9800 &        837& &       34\,513& &      512\,744& &        149\,997& &        1&300 &    0&566 &    0&356 &    0&390\\
     25& &      5&00 &    0&9131 &        521& &       45\,516& &      364\,680& &        184\,780& &        1&293 &    0&568 &    0&359 &    0&393\\
     30& &      0&05 &    1&4417 &     2\,251& &       10\,697& &   1\,031\,299& &         57\,276& &        1&302 &    0&566 &    0&355 &    0&395\\
     30& &      0&10 &    1&2210 &     1\,942& &       21\,054& &   1\,009\,122& &         94\,918& &        1&324 &    0&560 &    0&346 &    0&386\\
     30& &      0&50 &    1&0186 &     1\,178& &       44\,101& &      755\,593& &        164\,995& &        1&324 &    0&560 &    0&345 &    0&379\\
     30& &      1&00 &    0&9639 &        951& &       55\,015& &      657\,633& &        194\,894& &        1&321 &    0&560 &    0&347 &    0&380\\
     30& &      5&00 &    0&8894 &        587& &       75\,828& &      465\,898& &        248\,448& &        1&319 &    0&561 &    0&347 &    0&382\\
     35& &      0&05 &    1&4087 &     2\,355& &       16\,000& &   1\,248\,903& &         71\,920& &        1&315 &    0&562 &    0&349 &    0&391\\
     35& &      0&10 &    1&2392 &     2\,087& &       27\,668& &   1\,218\,128& &        108\,155& &        1&342 &    0&555 &    0&338 &    0&375\\
     35& &      0&50 &    1&0140 &     1\,338& &       62\,590& &      941\,772& &        199\,702& &        1&344 &    0&554 &    0&337 &    0&371\\
     35& &      1&00 &    0&9501 &     1\,089& &       81\,296& &      823\,309& &        243\,366& &        1&343 &    0&554 &    0&337 &    0&371\\
     35& &      5&00 &    0&8678 &        653& &      117\,258& &      573\,154& &        321\,847& &        1&346 &    0&554 &    0&336 &    0&372\\
     40& &      0&05 &    1&4201 &     2\,460& &       20\,626& &   1\,456\,790& &         81\,371& &        1&331 &    0&558 &    0&342 &    0&384\\
     40& &      0&10 &    1&2789 &     2\,220& &       32\,737& &   1\,421\,580& &        114\,919& &        1&359 &    0&550 &    0&330 &    0&365\\
     40& &      0&50 &    1&0115 &     1\,477& &       84\,607& &   1\,156\,557& &        234\,911& &        1&362 &    0&549 &    0&329 &    0&363\\
     40& &      1&00 &    0&9389 &     1\,211& &      114\,372& &   1\,016\,621& &        295\,137& &        1&364 &    0&548 &    0&328 &    0&363\\
     40& &      5&00 &    0&8479 &        721& &      173\,644& &      687\,696& &        406\,412& &        1&372 &    0&546 &    0&325 &    0&363\\
     45& &      0&05 &    1&4571 &     2\,550& &       23\,906& &   1\,644\,054& &         85\,812& &        1&344 &    0&554 &    0&337 &    0&378\\
     45& &      0&10 &    1&3266 &     2\,341& &       36\,613& &   1\,617\,554& &        118\,015& &        1&373 &    0&546 &    0&325 &    0&357\\
     45& &      0&50 &    1&0104 &     1\,587& &      109\,823& &   1\,394\,789& &        269\,944& &        1&379 &    0&544 &    0&322 &    0&356\\
     45& &      1&00 &    0&9293 &     1\,312& &      154\,381& &   1\,233\,338& &        349\,384& &        1&384 &    0&543 &    0&320 &    0&356\\
     45& &      5&00 &    0&8296 &        797& &      246\,867& &      810\,090& &        500\,992& &        1&397 &    0&539 &    0&314 &    0&355\\
     50& &      0&05 &    1&5084 &     2\,624& &       26\,059& &   1\,812\,465& &         86\,914& &        1&353 &    0&552 &    0&333 &    0&372\\
     50& &      0&10 &    1&3661 &     2\,443& &       40\,788& &   1\,816\,290& &        121\,580& &        1&382 &    0&544 &    0&321 &    0&352\\
     50& &      0&50 &    1&0105 &     1\,696& &      137\,864& &   1\,657\,050& &        304\,354& &        1&394 &    0&540 &    0&316 &    0&350\\
     50& &      1&00 &    0&9212 &     1\,416& &      201\,021& &   1\,473\,950& &        405\,043& &        1&401 &    0&538 &    0&313 &    0&350\\
     50& &      5&00 &    0&8133 &        877& &      337\,922& &      939\,025& &        603\,700& &        1&420 &    0&533 &    0&305 &    0&347\\
     55& &      0&05 &    1&5706 &     2\,692& &       27\,233& &   1\,961\,060& &         85\,721& &        1&362 &    0&549 &    0&329 &    0&366\\
     55& &      0&10 &    1&4046 &     2\,529& &       44\,642& &   2\,009\,304& &        124\,155& &        1&389 &    0&542 &    0&318 &    0&348\\
     55& &      0&50 &    1&0124 &     1\,807& &      168\,106& &   1\,939\,776& &        336\,623& &        1&406 &    0&537 &    0&311 &    0&346\\
     55& &      1&00 &    0&9152 &     1\,523& &      254\,050& &   1\,737\,899& &        460\,099& &        1&415 &    0&535 &    0&307 &    0&344\\
     55& &      5&00 &    0&7992 &        961& &      449\,023& &   1\,078\,605& &        712\,565& &        1&440 &    0&528 &    0&297 &    0&340\\
     60& &      0&05 &    1&6344 &     2\,755& &       28\,052& &   2\,098\,173& &         84\,004& &        1&371 &    0&547 &    0&325 &    0&361\\
     60& &      0&10 &    1&4394 &     2\,605& &       48\,560& &   2\,198\,748& &        126\,745& &        1&394 &    0&540 &    0&316 &    0&345\\
     60& &      0&50 &    1&0028 &     2\,022& &      210\,699& &   2\,180\,116& &        384\,420& &        1&412 &    0&535 &    0&308 &    0&342\\
     60& &      1&00 &    0&9044 &     1\,686& &      322\,274& &   1\,968\,043& &        528\,985& &        1&427 &    0&531 &    0&303 &    0&340\\
     60& &      5&00 &    0&7871 &     1\,046& &      576\,706& &   1\,239\,622& &        823\,831& &        1&455 &    0&524 &    0&291 &    0&334\\
     65& &      0&05 &    1&7083 &     2\,808& &       28\,001& &   2\,215\,973& &         80\,685& &        1&379 &    0&544 &    0&322 &    0&356\\
     65& &      0&10 &    1&4891 &     2\,670& &       50\,342& &   2\,363\,839& &        125\,232& &        1&400 &    0&538 &    0&313 &    0&341\\
     65& &      0&50 &    1&0113 &     2\,143& &      242\,171& &   2\,453\,178& &        410\,782& &        1&421 &    0&533 &    0&305 &    0&339\\
     65& &      1&00 &    0&9040 &     1\,835& &      384\,628& &   2\,244\,926& &        581\,941& &        1&438 &    0&528 &    0&298 &    0&336\\
     65& &      5&00 &    0&7792 &     1\,144& &      713\,179& &   1\,456\,962& &        926\,551& &        1&465 &    0&521 &    0&288 &    0&330\\
     70& &      0&05 &    1&9273 &     2\,830& &       21\,154& &   2\,211\,446& &         62\,872& &        1&407 &    0&537 &    0&310 &    0&334\\
     70& &      0&10 &    1&5424 &     2\,723& &       51\,330& &   2\,516\,482& &        122\,503& &        1&407 &    0&537 &    0&310 &    0&337\\
     70& &      0&50 &    1&0258 &     2\,247& &      267\,457& &   2\,738\,379& &        426\,572& &        1&426 &    0&531 &    0&303 &    0&336\\
     70& &      1&00 &    0&9137 &     1\,966& &      430\,915& &   2\,558\,262& &        610\,607& &        1&443 &    0&527 &    0&296 &    0&332\\
     70& &      5&00 &    0&7785 &     1\,275& &      830\,427& &   1\,812\,842& &        996\,941& &        1&466 &    0&521 &    0&287 &    0&328\\
     75& &      0&05 &    1&9854 &     2\,855& &       21\,758& &   2\,324\,345& &         62\,117& &        1&415 &    0&534 &    0&307 &    0&332\\
     75& &      0&10 &    1&5863 &     2\,769& &       53\,086& &   2\,670\,747& &        121\,465& &        1&412 &    0&535 &    0&309 &    0&334\\
     75& &      0&50 &    1&0459 &     2\,352& &      285\,273& &   3\,027\,221& &        432\,191& &        1&427 &    0&531 &    0&302 &    0&334\\
     75& &      1&00 &    0&9304 &     2\,108& &      461\,523& &   2\,903\,894& &        620\,803& &        1&443 &    0&527 &    0&296 &    0&331\\
     75& &      5&00 &    0&7962 &     1\,527& &      873\,708& &   2\,288\,897& &        998\,714& &        1&466 &    0&521 &    0&287 &    0&326\\
\hline\hline
\end{tabular}
\normalsize
\end{center}
\end{table}

\clearpage

\begin{table*}[p]
\begin{center}
\caption{Same as Table\,\ref{tab:cond},
for the most irradiated planets
($F_\star=4.18\times~10^6\,\mathrm{W.m^{-2}}$). An electronic version of this table is available at \textit{http://perso.ens-lyon.fr/jeremy.leconte/JLSite/JLsite/Exoplanets\_Simulations.html}}
\label{tab:irrad}
\small
\begin{tabular}{ r@{.}l r@{.}l r@{.}l r@{.}l r@{.}l r@{.}l r@{.}l  r@{.}l r@{.}l  r@{.}l | r@{.}l} \hline\hline 
\multicolumn{2}{c}{$\mp\,(\mjup)$}	&\multicolumn{2}{c}{Age (Gyr)} &\multicolumn{2}{c}{$R_\mathrm{0,p}\,(\rjup)$}  & \multicolumn{2}{c}{$\ind{T}{eff}\,(K)$} & \multicolumn{2}{c}{$\ind{P}{c}\,$(M\,Bar)} &\multicolumn{2}{c}{$\ind{T}{c}\,(K)$ }& \multicolumn{2}{c}{$\ind{\rho}{c}\,$(kg/m$^3$)} & \multicolumn{2}{c}{$n$} & \multicolumn{2}{c}{$\kappa_n$} & \multicolumn{4}{c}{$k_2$} \\
 \multicolumn{18}{c}{ } & \multicolumn{2}{c}{analytic}& \multicolumn{2}{c}{numeric} \\ \hline
  \hline
     0&5 &      0&05 &    1&4802 &       261&9 &           3&86 &       26\,591& &          1\,109& &        1&243 &    0&583 &    0&383 &    0&253\\
     0&5 &      0&10 &    1&3812 &       219&0 &           4&33 &       24\,369& &          1\,195& &        1&129 &    0&616 &    0&441 &    0&296\\
     0&5 &      0&50 &    1&2323 &       143&8 &           5&38 &       20\,269& &          1\,385& &        0&975 &    0&662 &    0&529 &    0&375\\
     0&5 &      1&00 &    1&1885 &       118&8 &           5&81 &       18\,784& &          1\,459& &        0&935 &    0&674 &    0&554 &    0&402\\
     0&5 &      5&00 &    1&1115 &        76&8 &           6&79 &       15\,221& &          1\,626& &        0&872 &    0&693 &    0&595 &    0&450\\
     0&6 &      0&05 &    1&4449 &       287&5 &           5&53 &       30\,336& &          1\,332& &        1&257 &    0&579 &    0&376 &    0&268\\
     0&6 &      0&10 &    1&3561 &       242&5 &           6&16 &       27\,836& &          1\,438& &        1&156 &    0&608 &    0&426 &    0&310\\
     0&6 &      0&50 &    1&2202 &       155&5 &           7&47 &       22\,690& &          1\,636& &        1&001 &    0&654 &    0&513 &    0&391\\
     0&6 &      1&00 &    1&1808 &       128&1 &           8&02 &       20\,870& &          1\,723& &        0&964 &    0&665 &    0&536 &    0&416\\
     0&6 &      5&00 &    1&1100 &        81&7 &           9&27 &       17\,153& &          1\,904& &        0&904 &    0&683 &    0&574 &    0&462\\
     0&7 &      0&05 &    1&4167 &       313&4 &           7&68 &       34\,787& &          1\,542& &        1&258 &    0&578 &    0&375 &    0&281\\
     0&7 &      0&10 &    1&3347 &       263&3 &           8&61 &       31\,797& &          1\,689& &        1&173 &    0&603 &    0&417 &    0&322\\
     0&7 &      0&50 &    1&2099 &       166&9 &          10&31 &       25\,479& &          1\,938& &        1&028 &    0&646 &    0&497 &    0&403\\
     0&7 &      1&00 &    1&1734 &       137&1 &          11&03 &       23\,351& &          2\,028& &        0&993 &    0&656 &    0&518 &    0&427\\
     0&7 &      5&00 &    1&1083 &        87&1 &          12&62 &       19\,183& &          2\,231& &        0&936 &    0&674 &    0&553 &    0&471\\
     0&8 &      0&05 &    1&3997 &       335&1 &           9&86 &       38\,796& &          1\,706& &        1&234 &    0&585 &    0&387 &    0&295\\
     0&8 &      0&10 &    1&3207 &       280&5 &          11&21 &       35\,538& &          1\,893& &        1&169 &    0&604 &    0&419 &    0&334\\
     0&8 &      0&50 &    1&2015 &       177&1 &          13&69 &       28\,328& &          2\,228& &        1&049 &    0&639 &    0&485 &    0&411\\
     0&8 &      1&00 &    1&1680 &       145&0 &          14&52 &       25\,811& &          2\,319& &        1&015 &    0&650 &    0&505 &    0&434\\
     0&8 &      5&00 &    1&1064 &        91&9 &          16&56 &       21\,024& &          2\,555& &        0&962 &    0&666 &    0&537 &    0&476\\
     0&9 &      0&05 &    1&3902 &       354&1 &          12&10 &       42\,282& &          1\,859& &        1&195 &    0&596 &    0&406 &    0&309\\
     0&9 &      0&10 &    1&3123 &       295&9 &          13&92 &       38\,890& &          2\,074& &        1&149 &    0&610 &    0&430 &    0&346\\
     0&9 &      0&50 &    1&1955 &       186&6 &          17&37 &       31\,105& &          2\,479& &        1&059 &    0&636 &    0&479 &    0&417\\
     0&9 &      1&00 &    1&1642 &       152&2 &          18&43 &       28\,174& &          2\,599& &        1&028 &    0&646 &    0&497 &    0&439\\
     0&9 &      5&00 &    1&1048 &        96&5 &          21&02 &       22\,830& &          2\,866& &        0&981 &    0&660 &    0&525 &    0&479\\
     1&0 &      0&05 &    1&3845 &       372&0 &          14&45 &       45\,609& &          2\,005& &        1&155 &    0&608 &    0&427 &    0&322\\
     1&0 &      0&10 &    1&3074 &       310&6 &          16&76 &       42\,025& &          2\,248& &        1&124 &    0&617 &    0&443 &    0&357\\
     1&0 &      0&50 &    1&1918 &       195&7 &          21&29 &       33\,727& &          2\,719& &        1&060 &    0&636 &    0&478 &    0&424\\
     1&0 &      1&00 &    1&1616 &       159&4 &          22&69 &       30\,526& &          2\,863& &        1&035 &    0&644 &    0&493 &    0&445\\
     1&0 &      5&00 &    1&1038 &       100&7 &          25&92 &       24\,599& &          3\,168& &        0&994 &    0&656 &    0&517 &    0&482\\
     1&2 &      0&05 &    1&3802 &       403&7 &          19&45 &       51\,893& &          2\,287& &        1&082 &    0&630 &    0&466 &    0&347\\
     1&2 &      0&10 &    1&3038 &       337&8 &          22&81 &       47\,931& &          2\,582& &        1&070 &    0&633 &    0&473 &    0&379\\
     1&2 &      0&50 &    1&1886 &       213&2 &          29&76 &       38\,662& &          3\,176& &        1&048 &    0&640 &    0&485 &    0&437\\
     1&2 &      1&00 &    1&1592 &       173&3 &          31&97 &       35\,060& &          3\,362& &        1&034 &    0&644 &    0&494 &    0&454\\
     1&2 &      5&00 &    1&1032 &       108&7 &          36&81 &       28\,050& &          3\,754& &        1&007 &    0&652 &    0&510 &    0&487\\
     1&5 &      0&05 &    1&3816 &       441&7 &          28&32 &       59\,797& &          2\,739& &        1&010 &    0&651 &    0&508 &    0&375\\
     1&5 &      0&10 &    1&3063 &       371&5 &          33&41 &       55\,635& &          3\,094& &        1&009 &    0&652 &    0&508 &    0&403\\
     1&5 &      0&50 &    1&1899 &       236&2 &          44&60 &       45\,509& &          3\,844& &        1&017 &    0&649 &    0&503 &    0&453\\
     1&5 &      1&00 &    1&1602 &       191&8 &          48&41 &       41\,202& &          4\,100& &        1&016 &    0&649 &    0&504 &    0&466\\
     1&5 &      5&00 &    1&1047 &       119&5 &          56&43 &       32\,981& &          4\,612& &        1&005 &    0&653 &    0&511 &    0&494\\
     1&8 &      0&05 &    1&3842 &       476&2 &          39&02 &       67\,811& &          3\,207& &        0&976 &    0&662 &    0&528 &    0&394\\
     1&8 &      0&10 &    1&3099 &       401&5 &          46&16 &       62\,943& &          3\,636& &        0&976 &    0&661 &    0&528 &    0&421\\
     1&8 &      0&50 &    1&1931 &       257&0 &          62&12 &       51\,794& &          4\,534& &        0&993 &    0&656 &    0&518 &    0&464\\
     1&8 &      1&00 &    1&1625 &       208&6 &          67&92 &       47\,144& &          4\,853& &        1&000 &    0&654 &    0&514 &    0&476\\
     1&8 &      5&00 &    1&1068 &       129&5 &          79&87 &       37\,666& &          5\,492& &        0&999 &    0&655 &    0&514 &    0&499\\
     2&1 &      0&05 &    1&3877 &       507&3 &          51&29 &       77\,544& &          3\,643& &        0&963 &    0&665 &    0&536 &    0&408\\
     2&1 &      0&10 &    1&3137 &       429&3 &          61&02 &       71\,611& &          4\,159& &        0&964 &    0&665 &    0&536 &    0&432\\
     2&1 &      0&50 &    1&1963 &       276&7 &          82&88 &       58\,167& &          5\,246& &        0&980 &    0&660 &    0&526 &    0&472\\
     2&1 &      1&00 &    1&1650 &       224&2 &          90&91 &       53\,093& &          5\,622& &        0&990 &    0&657 &    0&520 &    0&482\\
     2&1 &      5&00 &    1&1087 &       138&8 &         107&73 &       42\,307& &          6\,398& &        0&996 &    0&656 &    0&516 &    0&502\\
     3&0 &      0&05 &    1&3997 &       583&5 &          96&63 &      106\,857& &          4\,903& &        0&960 &    0&666 &    0&538 &    0&435\\
     3&0 &      0&10 &    1&3263 &       499&2 &         116&54 &       99\,725& &          5\,658& &        0&964 &    0&665 &    0&536 &    0&453\\
     3&0 &      0&50 &    1&2058 &       329&1 &         164&27 &       80\,267& &          7\,377& &        0&983 &    0&659 &    0&524 &    0&480\\
     3&0 &      1&00 &    1&1709 &       267&7 &         183&21 &       72\,047& &          8\,026& &        0&993 &    0&657 &    0&518 &    0&488\\
     3&0 &      5&00 &    1&1108 &       163&4 &         222&49 &       56\,621& &          9\,310& &        1&006 &    0&653 &    0&510 &    0&501\\
\hline\hline
\end{tabular}
\normalsize
\end{center}
\end{table*}
\clearpage

\begin{table}[p]
\begin{center}
\caption{End of Table\,\ref{tab:irrad}.}
\label{tab:irrad2}
\small
\begin{tabular}{ r@{.}l r@{.}l r@{.}l r@{.}l r@{.}l r@{.}l r@{.}l  r@{.}l r@{.}l  r@{.}l | r@{.}l} \hline\hline 
\multicolumn{2}{c}{$\mp\,(\mjup)$}	&\multicolumn{2}{c}{Age (Gyr)} &\multicolumn{2}{c}{$R_\mathrm{0,p}\,(\rjup)$}  & \multicolumn{2}{c}{$\ind{T}{eff}\,(K)$} & \multicolumn{2}{c}{$\ind{P}{c}\,$(M\,Bar)} &\multicolumn{2}{c}{$\ind{T}{c}\,(K)$ }& \multicolumn{2}{c}{$\ind{\rho}{c}\,$(kg/m$^3$)} & \multicolumn{2}{c}{$n$} & \multicolumn{2}{c}{$\kappa_n$} & \multicolumn{4}{c}{$k_2$} \\
 \multicolumn{18}{c}{ } & \multicolumn{2}{c}{analytic}& \multicolumn{2}{c}{numeric} \\ \hline
 \hline
     5&0 &      0&05 &    1&4224 &       741&1 &         247&80 &      166\,910& &          7\,870& &        0&997 &    0&655 &    0&515 &    0&457\\
     5&0 &      0&10 &    1&3473 &       619&5 &         304&77 &      157\,781& &          9\,229& &        1&007 &    0&652 &    0&509 &    0&466\\
     5&0 &      0&50 &    1&2133 &       419&3 &         464&43 &      128\,666& &         12\,786& &        1&044 &    0&641 &    0&488 &    0&476\\
     5&0 &      1&00 &    1&1708 &       344&7 &         537&24 &      115\,271& &         14\,293& &        1&059 &    0&637 &    0&479 &    0&478\\
     5&0 &      5&00 &    1&1014 &       213&8 &         684&92 &       90\,171& &         17\,174& &        1&077 &    0&631 &    0&469 &    0&483\\
     8&0 &      0&05 &    1&4420 &       932&1 &         636&38 &      259\,574& &         12\,943& &        1&086 &    0&628 &    0&464 &    0&449\\
     8&0 &      0&10 &    1&3495 &       793&0 &         831&43 &      248\,326& &         15\,866& &        1&105 &    0&623 &    0&453 &    0&453\\
     8&0 &      0&50 &    1&2051 &       525&5 &      1\,327&72 &      212\,307& &         22\,576& &        1&149 &    0&610 &    0&430 &    0&453\\
     8&0 &      1&00 &    1&1571 &       445&5 &      1\,570&48 &      192\,724& &         25\,593& &        1&164 &    0&605 &    0&422 &    0&451\\
     8&0 &      5&00 &    1&0748 &       278&7 &      2\,111&44 &      147\,177& &         31\,873& &        1&179 &    0&601 &    0&414 &    0&452\\
    10&0 &      0&05 &    1&4483 &    1\,051&8 &      1\,016&77 &      325\,164& &         16\,633& &        1&143 &    0&612 &    0&433 &    0&438\\
    10&0 &      0&10 &    1&3480 &       893&7 &      1\,359&72 &      313\,318& &         20\,717& &        1&163 &    0&606 &    0&422 &    0&440\\
    10&0 &      0&50 &    1&1931 &       588&7 &      2\,240&54 &      270\,490& &         30\,079& &        1&200 &    0&595 &    0&403 &    0&438\\
    10&0 &      1&00 &    1&1450 &       500&6 &      2\,647&31 &      247\,482& &         34\,031& &        1&210 &    0&592 &    0&399 &    0&437\\
    10&0 &      5&00 &    1&0583 &       321&2 &      3\,616&15 &      188\,199& &         42\,848& &        1&217 &    0&590 &    0&395 &    0&437\\
    13&0 &      0&05 &    1&4543 &    1\,227&3 &      1\,771&70 &      427\,626& &         22\,312& &        1&207 &    0&593 &    0&400 &    0&422\\
    13&0 &      0&10 &    1&3420 &    1\,036&9 &      2\,453&77 &      414\,870& &         28\,462& &        1&228 &    0&587 &    0&390 &    0&422\\
    13&0 &      0&50 &    1&1746 &       684&1 &      4\,190&16 &      360\,362& &         42\,333& &        1&249 &    0&581 &    0&380 &    0&421\\
    13&0 &      1&00 &    1&1260 &       572&1 &      4\,952&80 &      331\,095& &         47\,900& &        1&251 &    0&580 &    0&379 &    0&421\\
    13&0 &      5&00 &    1&0376 &       378&2 &      6\,812&26 &      253\,286& &         60\,644& &        1&245 &    0&582 &    0&381 &    0&422\\
    17&0 &      0&05 &    1&9236 &    1\,814&9 &      1\,092&21 &      530\,529& &         13\,540& &        1&206 &    0&593 &    0&401 &    0&392\\
    17&0 &      0&10 &    1&5737 &    1\,547&6 &      2\,343&85 &      565\,858& &         24\,210& &        1&250 &    0&581 &    0&379 &    0&402\\
    17&0 &      0&50 &    1&1635 &       839&0 &      7\,729&08 &      492\,879& &         58\,757& &        1&285 &    0&571 &    0&363 &    0&405\\
    17&0 &      1&00 &    1&1059 &       676&3 &      9\,432&93 &      451\,135& &         68\,127& &        1&280 &    0&572 &    0&365 &    0&406\\
    17&0 &      5&00 &    1&0153 &       451&7 &     13\,139&79 &      350\,018& &         87\,181& &        1&267 &    0&576 &    0&371 &    0&409\\
    20&0 &      0&05 &    1&7413 &    1\,796&7 &      2\,127&33 &      643\,444& &         20\,955& &        1&257 &    0&579 &    0&376 &    0&392\\
    20&0 &      0&10 &    1&4005 &    1\,444&2 &      5\,080&99 &      666\,543& &         40\,165& &        1&300 &    0&566 &    0&356 &    0&395\\
    20&0 &      0&50 &    1&1447 &       901&3 &     11\,241&96 &      577\,707& &         72\,520& &        1&299 &    0&567 &    0&357 &    0&398\\
    20&0 &      1&00 &    1&0891 &       731&1 &     13\,657&81 &      530\,282& &         83\,878& &        1&292 &    0&569 &    0&360 &    0&399\\
    20&0 &      5&00 &    1&0015 &       492&3 &     18\,915&42 &      418\,729& &        106\,994& &        1&277 &    0&573 &    0&366 &    0&402\\
\hline\hline
\end{tabular}
\normalsize
\end{center}
\end{table}



\end{document}